\documentclass[12pt]{article}
\usepackage{amssymb,bbm,amsthm,amscd,amsmath}
\usepackage{epsfig,graphics}
\usepackage{setspace}

\newcommand{\II}{{\mathbbm{1}}}
\theoremstyle{theorem}
\newtheorem{Theorem}{Theorem}
\newtheorem{Proposition}{Proposition}
\newtheorem{Lemma}{Lemma}

\newtheorem{Definition}{Definition}
\newtheorem{Punkt}{}[subsection]

\theoremstyle{remark}
\newtheorem{Remark}{Remark}

\newcommand{\btm}{\begin{Theorem}}
\newcommand{\etm}{\end{Theorem}}

\newcommand{\ben}{\begin{enumerate}}
\newcommand{\een}{\end{enumerate}}

\newcommand{\bpu}{\begin{Punkt}\rm}
\newcommand{\epu}{\end{Punkt}}

\newcommand{\bre}{\begin{Remark}\rm}
\newcommand{\ere}{\end{Remark}}

\newcommand{\ble}{\begin{Lemma}}
\newcommand{\ele}{\end{Lemma}}

\newcommand{\bpr}{\begin{Proposition}}
\newcommand{\epr}{\end{Proposition}}

\newcommand{\beq}{\begin{equation}}
\newcommand{\eeq}{\end{equation}}

\newcommand{\CC}{{\mathbb{C}}}

\newcommand{\RR}{{\mathbb{R}}}

\newcommand{\ZZ}{{\mathbb{Z}}}
\DeclareMathOperator{\Ad}{Ad}
\DeclareMathOperator{\diag}{diag}
\DeclareMathOperator{\Imag}{Im}
\DeclareMathOperator{\Pol}{Pol}
\DeclareMathOperator{\tr}{tr}
\newcommand{\sinfn}[1]{\sin(#1)}
\newcommand{\cosfn}[1]{\cos(#1)}
\newcommand{\arcsinfn}[1]{\arcsin(#1)}
\newcommand{\sinfnpot}[2]{\sin^{#1}(#2)}
\newcommand{\cosfnpot}[2]{\cos^{#1}(#2)}
\newcommand{\tanfn}[1]{\tan(#1)}
\newcommand{\AC}{\mr{AC}}
\newcommand{\cdic}{C^\infty_\mr{ev}[0,\pi]}
\newcommand{\mc}[1]{\mathcal{#1}}
\newcommand{\mf}[1]{\mathfrak{#1}}
\newcommand{\mr}[1]{\mathrm{#1}}
\newcommand{\OPg}{{\mr P}}
\newcommand{\Hi}{{\mc H}}
\newcommand{\scale}{\beta}

\newcommand{\comment}[1]{}
\newcommand{\verweis}[1]{}
\newcommand{\todo}[1]{}
\newcommand{\ddx}{\frac{\mr d}{\mr d x}}
\newcommand{\ddxx}{\frac{\mr d^2}{\mr d x^2}}
\newcommand{\ddt}{\frac{\mr d}{\mr d t}}
\newcommand{\ddtn}{\left.\frac{\mr d}{\mr d t}\right|_{t=0}}
\newcommand{\ddsn}{\left.\frac{\mr d}{\mr d s}\right|_{s=0}}
\newcommand{\SU}{\mr{SU}}

\newcommand{\SL}{\mr{SL}}
\newcommand{\inco}{\nu}
\newcommand{\group}{G}
\newcommand{\lieal}{{\mf g}}
\newcommand{\cfg}{{\mc X}}
\newcommand{\pha}{{\mc P}}
\newcommand{\vol}{\mr{vol}}
\newcommand{\Ann}{\mr{Ann}}

\newcommand{\braket}[2]{\langle#1|#2\rangle}

\newcommand{\vp}{\varphi}

\newcommand{\ctg}{\mr T^\ast}
\newcommand{\tg}{\mr T}
\newcommand{\rref}[1]{{\rm \ref{#1}}}
\newcommand{\cl}{\mr{c}}
\newcommand{\ol}[1]{\overline{#1}}

\newcommand{\linie}[3]{\put(#1){\line(#2){#3}}}
\newcommand{\marke}[3]{\put(#1){\put(0.05,0.1){\makebox(-0.1,-0.2)[#2]{$#3$}}}}

\begin{document}

\title{\bf On the algebra of quantum observables for a certain
gauge model}

\author{
    G.~Rudolph and M.~Schmidt\\
    Institut f\"ur Theoretische Physik, Universit\"at Leipzig\\
    Augustusplatz 10/11, 04109 Leipzig, Germany\\
    }

\maketitle

\comment{
\vspace{2cm}

{\bf Keywords:}~
\\

{\bf MSC:}~ 70G65, 70S15
}
\vspace{2cm}

\begin{abstract}

\noindent
We prove that the algebra of observables of a certain gauge model is generated
by unbounded elements in the sense of  Woronowicz. The generators are
constructed from the classical generators of invariant polynomials by means of
geometric quantization.

\end{abstract}

\newpage



\section{Introduction}
\label{S-intro}


One of the fundamental structures in nonperturbative quantum field theory is the algebra of
observables and its representations. To construct the observable algebra and to find its
irreducible representations for a gauge theory is a complicated task, see
\cite{StrocchiWightman}, \cite{Fredenhagen} and \cite{Froehlich} for attempts
made in the seventies and eighties. Roughly speaking, one has to start with a
model of the field algebra carrying the action of the gauge group by
automorphisms, next one has to pass to the algebra of gauge invariant elements
and, finally, one has to factorize this algebra by an ideal generated by the
Gauss law. Unfortunately, standard charge superselection theory
\cite{DHR71,DHR74,DR90} does
not apply to genuine local gauge theories, see \cite{Buchholz82,Buchholz86}.

In order to separate functional analytical problems related to the mathematical
nature of quantum fields on continuous space time from those related to the
gauge structure, one is tempted to consider, in a first step, models
approximated on a finite lattice. In this context, we have constructed the
observable algebras and classified their irreducible representations both for
quantum electrodynamics \cite{KiRuTh,KiRuSl} and for quantum chromodynamics
\cite{KiRu02,KiRu05}. An additional challenge comes from the fact that on the
classical level there are nongeneric gauge orbit strata, see \cite{rsv:review}
for a review, which should have an impact on quantum level as well. In
\cite{hrs} we have shown that one can include these singularities by using the
concept of a costratified Hilbert space \cite{Hue:qr}. In the case of
chromodynamics, a full understanding of the observable algebra in terms of
generators and defining relations is still lacking, see \cite{JaKiRu} for
preliminary results. Generally speaking, gauge invariant generators are
polynomial invariants built from gauge and matter fields, corresponding to
classical generators of the algebra of polynomial invariants. Since typical
quantum observables are unbounded operators, one cannot hope to incorporate all
observables in a na\"ive sense into the observable algebra. Fortunately, there is
a suitable approach developed by Woronowicz in the nineties \cite{Woro}, which
makes it possible to say that a  given number of unbounded elements generates a
certain $C^*$-algebra, with the generators being affiliated with the algebra
under consideration in the $C^*$-sense. We remark that recently another
construction of a $C^\ast$-algebra of observables from unbounded physical
quantities was invented, see \cite{BuchholzGrundling}. In \cite{KiRu05} we have
shown that the field algebra of quantum chromodynamics is a $C^\ast$-algebra of
this type. In the present paper we prove that the algebra of observables of the
model studied in \cite{hrs} is also generated by unbounded operators in the
sense of Woronowicz. It is a challenge to extend this result to full chromodynamics on
a finite lattice in the future. In the case at hand, the generating operators
are the quantum counterparts of the generators of the algebra of real invariant
polynomials on the reduced phase space. This is an interesting fact in itself,
because in the Woronowicz theory there does not exist a general method to find a set of
generators of a given  $C^\ast$-algebra, nor does there exist a general method to find
the $C^\ast$-algebra generated by a given set of unbounded operators.

The paper is organized as follows: In Section \ref{S-model} we briefly present
the underlying classical model. In Section \ref{S-cobs} we present the algebra
of classical observables and its generators. Section \ref{S-qobs} is devoted to quantum observables. First
we quantize the classical generators using geometric quantization. Next, we discuss
the spectral properties of the quantized generators and the quantum
counterpart of the relation amongst the classical generators. Then, 
we construct the algebra of quantum observables and discuss the relations between our generators and the generators 
defined in \cite{KiRu05}. Finally, we comment on quantum dynamics and give an outlook.


\section{The model}
\label{S-model}

\begin{figure}

\unitlength1cm

 \begin{picture}(6,7.8)

\put(0.5,4.25){
 \put(1,0.27){
 \marke{4,1.82}{cl}{~-\!\II}
 \marke{0,1.82}{cr}{+\II~~}
 \marke{2,0}{tc}{T}
 \put(-0.135,1.82){\circle*{0.2}}
 \put(3.985,1.82){\circle*{0.2}}
 }
 \put(-1,0){
 \put(0,0.5){\epsfig{file=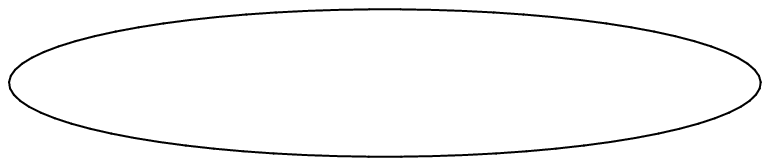,width=6cm,height=2cm}}
 }
} \put(0.5,1){
 \put(1,0.27){
 \marke{0,0}{tr}{\cfg_+~}
 \marke{2,0}{bc}{\cfg_0}
 \marke{4,0}{tl}{~\cfg_-}
 \marke{2,-0.5}{tc}{\cfg \cong T/W}
 \put(0,-0.06){\circle*{0.2}}
 \put(4,-0.06){\circle*{0.2}}
 \linie{0,-0.06}{1,0}{4}
 }
} \put(8.5,4.25){
 \put(1,0.27){
 \marke{4,1.82}{cl}{~(-\II,0)}
 \marke{0,1.82}{cr}{(+\II,0)~~}
 \marke{2,0}{tc}{T\times\mf t}
 \put(-0.135,1.82){\circle*{0.2}}
 \put(3.985,1.82){\circle*{0.2}}
 }
 \put(-1,0){
 \put(1.85,0.55){\epsfig{file=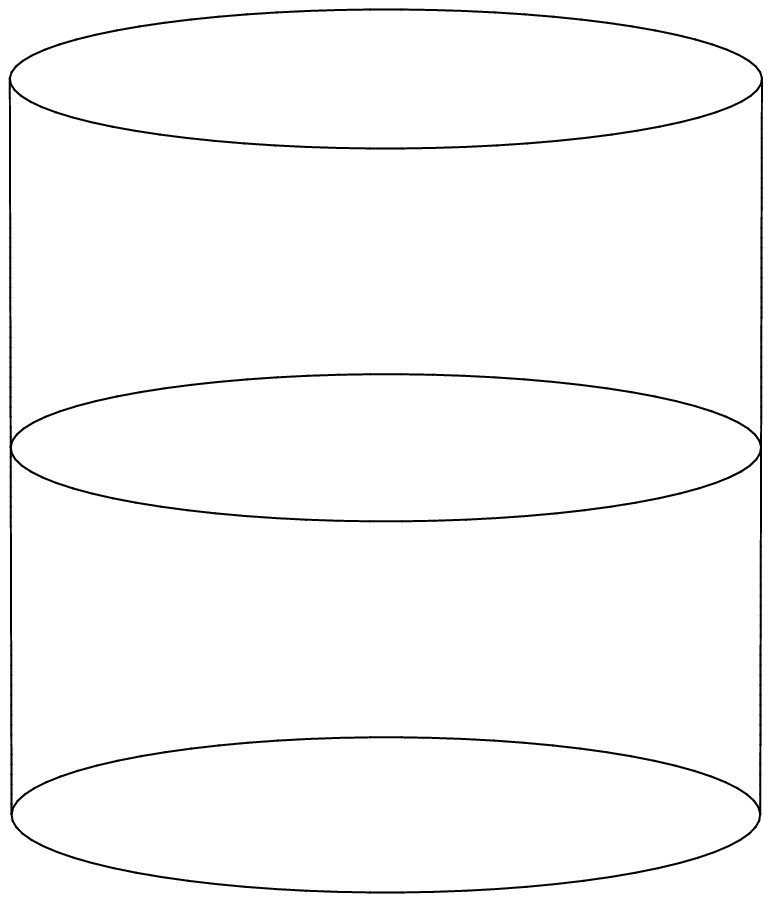,width=4.15cm,height=3cm}}
 }
} \put(8.5,1){
 \put(1,0.27){
 \marke{0,0}{tr}{\pha_+}
 \marke{4,0}{tl}{~\pha_-}
 \marke{4,1.75}{tl}{~\pha_0}
 \marke{2,-0.5}{tc}{\pha = \big(T\times\mf t\big)/W}
 \put(0,-0.06){\circle*{0.2}}
 \put(4,-0.06){\circle*{0.2}}
 }
 \put(-1,0){
 \put(0,0.02){\epsfig{file=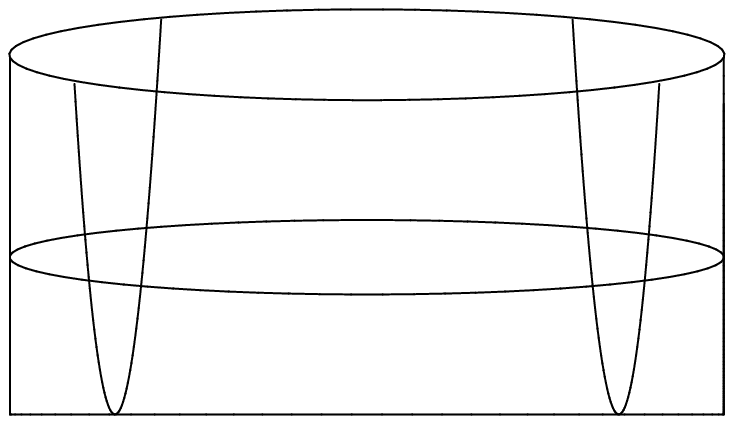,width=6cm,height=2cm}}
 }
}
 \end{picture}
\caption{\label{Fig-canoe}
Reduced configuration space $\cfg$ and reduced phase space
$\pha$ of the model together with their stratification by connected components
of orbit types}

 \end{figure}

The model was explained in detail in \cite{hrs}. We recall the main facts.
The configuration space is the group manifold $G = \SU(2)$, acted upon by $G$
itself by inner automorphisms,
$$
g\cdot a = g a g^{-1}\,.
$$
The phase space is given by the cotangent bundle $\ctg G$, acted upon by the
lifted action. This action is symplectic and it possesses a natural equivariant
momentum mapping $\mu : \ctg G\to\mf g^\ast$, where $\mf g$ denotes the Lie
algebra of $G$. Thus, the phase space carries the structure of a Hamiltonian
$G$-manifold. We trivialize $\ctg G\cong G\times\mf g$ by means of an invariant
scalar product $\langle\cdot,\cdot\rangle$ on $\mf g$ and left translation. In
these coordinates, the lifted
action is given by
$$
g\cdot(a,X)
 =
(gag^{-1},\Ad(g)X)
 \,,~~~~~~
a\in G, X\in \mf g, g\in G\,,
$$
and the natural momentum mapping is given by
 \beq\label{G-momap}
\mu(a,X) = aXa^{-1} - X\,.
 \eeq
W.r.t.\ the natural decomposition
 \beq\label{G-decotg}
\tg_{(a,X)} (G\times\mf g)
 =
\tg_a G \oplus \tg_X\mf g\,,
 \eeq
tangent vectors at $(a,X)\in G\times \mf g$ can be written in the form
 \beq\label{G-vf}
(\mr L_a' A, B)\,,
 \eeq
where $A,B\in\mf g$ and $\mr L_a$ means left multiplication by $a$. In this
notation, the symplectic potential reads
 \beq\label{G-splpot}
\theta_{(a,X)}\big((\mr L_a' A,B)\big) = \langle X,A \rangle
 \,,~~~~~~
A,B\in\mf g\,,
 \eeq
and the symplectic form $\omega = -\mr d\theta$ is given by
 \beq\label{G-omega}
\omega_{(a,X)}\big((\mr L_a' A_1,B_1)\,,\,(\mr L_a' A_2,B_2)\big)
 =
\langle A_1,B_2 \rangle
 -
\langle A_2,B_1 \rangle
 +
\langle X,[A_1,A_2]\rangle\,.
 \eeq
The model can be interpreted as an $\SU(2)$-lattice gauge theory on a single
spatial plaquette in the Hamiltonian approach in the tree gauge,
or as $\SU(2)$-gauge theory on a space-time cylinder in the temporal gauge and
after reduction by the group of based gauge transformations, see
\cite{hrs}. In both cases, the classical Hamiltonian is given by
 \beq\label{G-Ham}
 \textstyle
H(a,X)
 =
 -
\frac{1}{2} |X|^2
 +
\frac{\inco}{2}
 \left(3 - \Re\,\tr(a)\right), \ a\in \group,\, X \in \lieal\,.
 \eeq
Let $T$ denote the subgroup of $G$ of diagonal matrices and $\mf t$ the
subalgebra of $\mf g$ of diagonal matrices. Let $W$ denote the Weyl group. It
acts on $T$ and $\mf t$ by permutation of entries. The reduced configuration
space $\cfg$ is given by the adjoint quotient
$$
\cfg = G/\Ad(G) \cong T/W\,.
$$
For general $\SU(n)$, this is an $(n-1)$-simplex.
For $\SU(2)$, the parameterization
 \beq\label{G-paramT}
\phi:\RR\to T
 \,,~~~~~~
x\mapsto\diag(\mr e^{\mr i x},\mr e^{-\mr i x})
 \eeq
induces a homeomorphism $[0,\pi]\cong\cfg$.
The reduced phase space is the zero level singular symplectic quotient
$$
\pha = \mu^{-1}(0)/G\,.
$$
Since, according to \eqref{G-momap}, $\mu(a,X)=0$ means that $a$
and $X$ commute and hence can be simultaneously diagonalized, $\pha$ may be
identified with the quotient $(T\times\mf t)/W$. For $\SU(2)$, this amounts to
the cylinder $\mr U(1)\times\RR$, factorized by reflection about the (virtual)
line connecting the points $(1,0)$ and $(-1,0)$, see Figure
\rref{Fig-canoe}. The space arising this way is known as the canoe. It coincides
with the phase space of a spherical pendulum, reduced at zero angular momentum
by the rotations about the vertical axis.

The reduced configuration space and the reduced phase space are
stratified by connected components of orbit type subsets,
$$
\cfg = \cfg_0 \cup \cfg_+ \cup \cfg_-
 \,,~~~~~~
\pha = \pha_0 \cup \pha_+ \cup \pha_-\,,
$$
where $\cfg_\pm$ consists of the class of $\pm\II$ and $\pha_\pm$ consists of
the class of the zero covector over $\pm\II$, see Figure \rref{Fig-canoe}.

\section{Classical observables}
\label{S-cobs}


The algebra of classical observables, as provided by standard singular
symplectic reduction at level $0$, is given by the quotient Poisson algebra
$$
\mc O_\cl = C^\infty(\ctg G)^G/V^G\,,
$$
where $V$ denotes the vanishing ideal of the closed subset $\mu^{-1}(0)$
\cite{ACG}. This algebra contains as a Poisson subalgebra the quotient
$\Pol(\ctg G)^G/V_{\mr{Pol}}^G$, where $\Pol(\ctg G)$
denotes the algebra of real polynomials on $\ctg G$ and $V_{\mr{Pol}}$
is the vanishing ideal of $\mu^{-1}(0)$ in this algebra. By definition, a
function on $\ctg G$ is polynomial if via the diffeomorphism $\ctg G\cong
G\times \mf g$ it corresponds to a function that is polynomial in the matrix
entries.

\bre\label{Rem-pnmalg}

One could also define polynomial functions on $\ctg G$ to be functions which via
the diffeomorphism $\ctg G \cong G^\CC$ correspond to elements of
$\Pol(G^\CC)$, i.e., to the functions on $G^\CC$ that are polynomial in the
matrix entries. This type of polynomial functions was used in
\cite{Hue:bedlewo}. Since polar decomposition is non-polynomial,
the two types of polynomial functions lead to completely different subalgebras
of $\mc O_\cl$ which intersect only in the constants.

\ere

The generators of $\Pol(\ctg G)^G$ are provided by invariant theory. For
$G=\SU(n)$ it is known that, via the diffeomorphism $\ctg G\cong G\times\mf g$,
a set of generators is provided by the real and imaginary parts of arbitrary
trace monomials of order $2^n-1$ in $a,a^\dagger \in G$ and $X\in\mf t$
\cite{Weyl}. By means of the fundamental trace identity and the Cayley-Hamilton
theorem this set of generators can be reduced considerably. For $\SU(2)$ there
remain 3 generators,
 \beq\label{G-generators}
 \textstyle
f_0(a,X) = \tr(a)
 \,,~~~~~~
f_1(a,X) = \frac{1}{2\beta^2} \tr(aX)
 \,,~~~~~~
f_2(a,X) = - \frac{1}{2\beta^2} \tr(X^2)
 \,.
 \eeq
Here $\beta$ is a scaling factor, defined by
$$
 \textstyle
\langle X,Y\rangle = -\frac{1}{2\beta^2} \tr(XY)
 \,,~~~~~~
X,Y\in\mf g\,.
$$
The functions $f_0$, $f_1$, $f_2$ are already real. For convenience, the
generators $f_1$ and $f_2$ have been rescaled by the scaling factor of the
invariant scalar product on $\mf g$. This way, $f_2$ is twice the kinetic
energy. In terms of the generators, the Hamiltonian \eqref{G-Ham} reads
 \beq\label{G-Ham-ivr}
 \textstyle
H = \frac{1}{2} f_2 + \frac{1}{2 g^2} (3 - f_0)\,.
 \eeq
I.e., up to a shift and up to a coupling parameter, $f_0$ is the potential
energy of the system.

\bre

For $G=\SU(n)$, $n\geq 3$, to cut to size the set of generators one
also has to make use of the fact that in the level set $\mu^{-1}(0)$, $a$ and
$X$ commute. This is not necessary for $\SU(2)$ though. I.e., here the set of
generators of invariant polynomials for the reduced phase space $\pha$ and for
the full quotient $\ctg G/G$ coincide.

\ere

\begin{figure}

\centering

\unitlength1cm

 \comment{
 \begin{picture}(12,4)
 \put(5.4,0.3){
\put(0,0){\vector(1,0){4}}
\put(0,0){\vector(0,1){4}}
\put(0,0){\vector(2,1){4}}
 \marke{4,0}{cl}{f_0}
 \marke{0,4}{bc}{f_2}
 \marke{4,2}{cl}{f_1}
 }
 }

 \begin{picture}(12,4.5)
\put(2.5,0.5){\epsfig{file=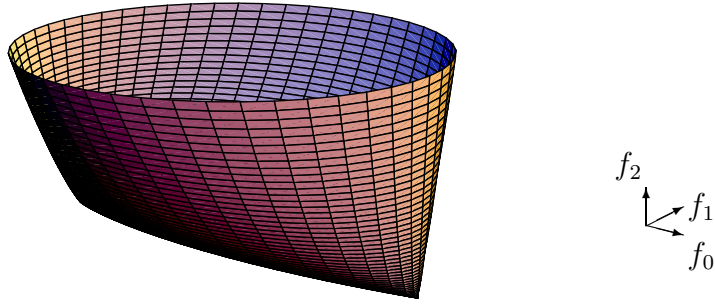,width=6cm,height=4cm}}
 \put(11,1.5){
\put(0,0){\vector(0,1){0.5}}
\put(0,0){\vector(4,-1){0.5}}
\put(0,0){\vector(2,1){0.5}}
 \marke{0,0.5}{br}{f_2}
 \marke{0.5,-0.125}{tl}{f_0}
 \marke{0.5,0.25}{cl}{f_1}
 }

 \end{picture}

\caption{\label{Fig-Himap}Image of $\pha$ under the Hilbert map defined by the generators
$f_0,f_1,f_2$. Equivalently, the semialgebraic subset of $\RR^3$ defined by
\eqref{G-rel} and \eqref{G-ineq}}

 \end{figure}

The generators $f_0$, $f_1$, $f_2$ define a map $\pha\to\RR^3$, known as the
Hilbert map associated with this set of generators. It is common knowledge, see
e.g.\ \cite{Schwarz}, that the Hilbert map is a homeomorphism onto its image and
that the image is a semialgebraic subset, i.e., a subset defined by equalities
and inequalities. The defining equalities and inequalities for our case are
obtained as follows. Up to diagonal conjugation, an arbitrary element $(a,X)\in
G\times\mf g$ can be written
$$
a = \left[\begin{array}{cc} \alpha & 0 \\ 0 & \ol{\alpha} \end{array}\right]
 \,,~~~~~~
X = \left[\begin{array}{cc} \mr i x & z \\ -\ol z & -\mr i x \end{array}\right]
 \,,~~~~~~
\alpha\in\mr U(1)\,,~~x\in\RR\,,~~z\in\CC\,.
$$
Then
$$
 \textstyle
f_0(a,X) = 2\Re(\alpha)
 \,,~~~~~~
f_1(a,X) = - \frac{1}{\beta^2} x \Imag(\alpha)
 \,,~~~~~~
f_2(a,X) = \frac{1}{\beta^2}(x^2 + |z|^2)\,.
$$
Eliminating $x$ and $\alpha$ we obtain the relations
 \beq\label{G-fullquotient}
(\scale^2 f_2 - |z|^2)(4 - f_0^2) - 4 \beta^4 f_1^2 = 0
 \,,~~~~~~
\scale^2 f_2 - |z|^2 \geq 0\,,
 \eeq
where now $f_0$, $f_1$ and $f_2$ are interpreted as standard coordinates in
$\RR^3$. Hence, the image of the full quotient $\ctg G/G$ under the Hilbert map
coincides with the set of points $(f_0,f_1,f_2)\in\RR^3$ satisfying
\eqref{G-fullquotient} for some $z$. Since \eqref{G-fullquotient} implies
$4-f_1^2 \geq 0$ and since $|z|$ can take any nonnegative value, this subset is
given by the two inequalities
$$
f_2 (4 - f_0^2) - 4 \scale^2 f_1^2 \geq 0
 \,,~~~~~~
4 - f_0^2 \geq 0\,.
$$
If $a$ and $X$ commute then, up to diagonal conjugation, $z=0$.
Hence, the reduced phase space $\pha\subseteq\ctg G/G$ corresponds to the subset
of \eqref{G-fullquotient} defined by $z=0$. Thus, this subset is given by the
relation
 \beq\label{G-rel}
f_2 (4 - f_0^2) - 4 \scale^2 f_1^2 = 0
 \eeq
and the inequality
 \beq\label{G-ineq}
f_2 \geq 0\,.
 \eeq
This subset is shown in Figure \rref{Fig-Himap}. It is of course a
concrete realization of the canoe, see Figure \rref{Fig-canoe}.
The image of the full quotient $\ctg G/G$ corresponds to this
subset together with the interior.

\bre

From Figure \rref{Fig-Himap} it is obvious that, topologically, $\pha$ is just a
copy of $\RR^2$. In fact, the Hilbert map defined by the natural generators of
the polynomial algebra $\Pol(G^\CC)^G$, see Remark \rref{Rem-pnmalg}, identifies
$\pha$ with the complex plane. However, as Poisson spaces, $\pha$ and $\CC$ are
distinct. This generalizes to $\SU(n)$. See \cite{Hue:bedlewo} for details.

\ere

Next, we compute the Hamiltonian vector fields $X_{f_i}$ associated with the
generators $f_i$ and the Poisson brackets between the generators. Let
$\OPg:\mr M_2(\CC)\to \mf g$ denote the orthogonal projection, i.e.,
 \beq\label{G-defP}
 \textstyle
\OPg(A) = \frac12(A-A^\dagger) - \frac{\mr i}{2} (\Imag\tr A) \II\,.
 \eeq
The defining equation for $X_{f_i}$ is
$
\omega(X_{f_i},Y) = - Y(f_i)
$
for all vector fields $Y$ on $G\times\mf g$. Writing
 \beq\label{G-Havf-AB}
(X_{f_i})_{(a,X)} = (\mr L_a' A_i,B_i)
 \eeq
and $Y_{(a,X)} = (\mr L_a' C,D)$ and using \eqref{G-omega} we obtain
$$
 \textstyle
\langle A_i,D \rangle - \langle B_i + [A_i,X],C \rangle
 =
- \ddtn f_i\left( a \mr e^{Ct} , X + t D \right)
 ~~~~~~
\forall~C,D\in\mf g\,.
$$
Evaluating the r.h.s.\ and solving for $A_i$ and $B_i$ we arrive at
 \begin{align}\label{G-Havf0}
A_0 & = 0 \,, & B_0 & = -2\scale^2 \OPg(a)\,,
\\ \label{G-Havf1}
A_1 & = \OPg(a)\,, & B_1 & = -\OPg(aX)\,,
\\ \label{G-Havf2}
A_2 & = -2X \,, & B_2 & = 0 \, .
 \end{align}
(Calculations are simplified by observing that $\tr(a)$ and $\tr(aX)$ are real,
hence the trace term in \eqref{G-defP} is absent in both cases.)

\ble\label{L-Havf}

The Hamiltonian vector fields $X_{f_0}$, $X_{f_1}$ and $X_{f_2}$ are
complete.

\ele

{\it Proof.}~ The flows of $X_{f_0}$ and $X_{f_2}$ are immediate:
$$
\Phi^{X_{f_0}}_t (a,X) = (a,X - 2\beta^2 \OPg(a) t)
 \,,~~~~~~
\Phi^{X_{f_2}}_t (a,X) = (a \mr e^{-2Xt},X)\,.
$$
They are defined for all $t\in\RR$. The flow equations for $X_{f_1}$ are
$$
\dot a = \mr L_a'\OPg(a)
 \,,~~~~~~
\dot X = -\OPg(aX)\,.
$$
Let $(a(0),X(0))$ be arbitrary but fixed initial values. Since $\dot a$ does
not depend on $X$ and since $G$ is compact, the solution $a(t)$ exists for all
$t\in\RR$. Hence, for $X_{f_1}$ to be complete it suffices that $|X(t)|$ be
finite for any $t\in\RR$. Consider the function $f(t) = |X(t)|^2$.
A brief computation using the Cayley-Hamilton theorem for $X$ yields $\ddt f(t)=
- \tr(a(t)) f(t)$. Then $\ddt f(t) \leq |\ddt f(t)| \leq 2 f(t)$. Thus, $f(t)$
is a nonnegative function whose derivative at $t$ is bounded by $2f(t)$. It
follows $f(t) \leq f(0) \mr e^{2t}$, hence the assertion.
 \qed
 \\

Finally, we calculate the Poisson brackets between the generators,
$$
 \textstyle
\{f_i,f_j\} = X_{f_i} f_j = \ddtn f_j(a\mr e^{A_it},X + tB_i)\,.
$$
Using the Cayley-Hamilton theorem to reduce powers of $a$ and $X$ we obtain
 \beq\label{G-Poibra}
 \textstyle
\{f_0,f_1\} = 2 - \frac12 f_0^2 \verweis{(SOA-16)}
 \,,~~~~~~
\{f_0,f_2\} = 4 \scale^2 f_1  \verweis{(SOA-16)}
 \,,~~~~~~
\{f_1,f_2\} = - f_0 f_2\,.  \verweis{(SOA-16)}
 \eeq
Since restriction of $f_i$ to the singular strata $\pha_\pm$ yields
$f_0|_{\pha_\pm} = \pm 2$ and $f_1|_{\pha_\pm} = f_2|_{\pha_\pm} = 0$,
$$
\{f_i,f_j\} |_{\pha_\pm} = 0\,.
$$
Thus, the Poisson structure of the reduced phase space $\pha$, given by
\eqref{G-Poibra}, reduces consistently to the (necessarily trivial) Poisson
structure on the singular strata $\pha_\pm$.


\section{Quantum observables}
\label{S-qobs}



\subsection{Quantization of classical generators}


The algebra of quantum observables will be constructed as follows. We quantize
the generators $f_i$ of the algebra of classical observables by means of
geometric quantization in the vertical polarization ('Schr\"odinger
quantization') on the unreduced phase space and subsequent reduction. As the
algebra
of quantum observables we will then take the $C^\ast$-algebra generated by these
operators in the sense of Woronowicz. The latter notion will be explained below.
We will loosely speak of the quantized generators as quantum observables as
well, although they do not belong to the algebra of quantum observables so
constructed.

The Hilbert space of Schr\"odinger quantization on $\ctg G$ can be identified
canonically with $L^2(G)$ with scalar product
$$
\braket{\psi_1}{\psi_2} = \frac{1}{\vol(G)} \int_G \ol{\psi_1}\psi_2 \mr
d a\,.
$$
Here $\mr d a$ stands for the volume form associated with the
bi-invariant Riemannian metric on $G$ defined by the invariant
scalar product on $\mf g$. By virtue of the isomorphism $G\times\mf g
\cong \tg G$, $f_2$ corresponds to the bi-invariant Riemannian metric defined
by the invariant scalar product on $\mf g$. Let $\Delta_G$ denote the Laplacian
associated with this metric. Since $G$ is closed, $\Delta_G$ is essentially
self-adjoint on the domain $C^\infty(G)$. The quantum observable $\hat f_2$
associated with $f_2$ is the unique self-adjoint extension of $- \hbar^2
\Delta_G$ see \cite[\S 9.7]{Woodhouse}. Thus, on the core $C^\infty(G)$,
 \beq\label{G-qobs2}
\hat f_2\psi = - \hbar^2 \Delta_G \psi
 \,,~~~~~~\psi\in C^\infty(G)\,.
 \eeq
In order to determine the quantum observables $\hat
f_0$ and $\hat f_1$ associated with the generators $f_0$ and $f_1$,
respectively, we have to recall the main steps in the construction of the
Hilbert space and the quantum observables in the Schr\"odinger quantization
\cite{Hall:Compact,Woodhouse}. Prequantization renders the complex line bundle
$L=\ctg G \times\CC$ with Hermitian form $h\big((a,z_1),(a,z_2)\big) = \ol{z_1}
z_2$ and connection $\nabla = \mr d + \theta \, , $ where $\theta$ denotes the symplectic potential 
of $\ctg G\, .$  Let $\pi:\ctg G \to G$ denote the canonical projection.
The vertical polarization is given by the vertical distribution $D\subseteq
\tg(\ctg G)$ induced by the fibres of $\ctg G$.
 \bigskip

{\it Hilbert space:}~ Consider the tautological complex line bundle $\kappa :=
\Lambda^n\Ann(D^\CC)$, where $n = \dim(G)$ and $\Ann$ denotes the annihilator of
$D^\CC$ in $(\tg^\CC)^\ast(\ctg G)$. The pull-back $\pi^\ast v$ of the volume
form $v$ associated with the Riemannian metric on $G$ defines a global section
in $\kappa$. Hence $\kappa$ is trivial and there exists a real line
bundle $\delta$ over $\ctg G$ such that $\kappa := (\delta\otimes\delta)^\CC$.
The bundle $\delta$ is called the half-form bundle associated with $D$. By
choosing a square root $\sqrt{\pi^\ast v}$ of $\pi^\ast v$ one obtains a global
nowhere vanishing section in $\delta$, hence $\delta$ is trivial, too.
Let $\Gamma_\mr{pol}(L\otimes\delta)$ denote the space of polarized sections in
$L\otimes\delta$. By definition, a section $\vp\otimes\nu$ in $L\otimes \delta$
is polarized if so are $\vp$ and $\nu$. A section $\vp\in \Gamma(L)$, viewed as
a function on $\ctg G$, is polarized if it is constant along the fibres, i.e.,
if $\vp = \pi^\ast \psi$ for some $\psi\in C^\infty(G)$. A section $\nu$ in
$\delta$ is polarized if $\nu\otimes\nu = \pi^\ast\alpha$ for some $n$-form
$\alpha$ on $G$. The Hilbert space $L^2_\mr{pol}(L\otimes\delta)$
is defined as the completion of $\Gamma_\mr{pol}(L\otimes\delta)$ w.r.t.\ the
norm defined by the following intrinsic scalar product:~ if $\vp_1\otimes\nu_1$
and $\vp_2\otimes\nu_2$ are polarized, $h(\vp_1,\vp_2)\nu_1\otimes\nu_2=
\pi^\ast\beta$ for some $n$-form $\beta$ on $G$. Then
$$
\braket{\vp_1\otimes\nu_1}{\vp_2\otimes\nu_2}
 :=
\frac{1}{\vol(G)} \int_G \beta\,.
$$
Finally, one can pass from half forms to functions on $G$ by observing that any
element of $\Gamma_\mr{pol}(L\otimes\delta)$ can be written in the form
$\vp\otimes\sqrt{\pi^\ast v}$ with $\vp = \pi^\ast \psi$ for some
$\psi\in C^\infty(G)$. By construction,
$$
\braket{\pi^\ast\psi_1\otimes\sqrt{\pi^\ast v}
 }{
\pi^\ast\psi_2\otimes\sqrt{\pi^\ast v}\rangle
 }
 =
\braket{\psi_1}{\psi_2}\,.
$$
Hence, the assignment
 \beq\label{G-ismSrqiz}
\psi\mapsto \pi^\ast\psi\otimes\sqrt{\pi^\ast v}
 \,,~~~~~~
\psi\in C^\infty(G)\,,
 \eeq
defines a unitary isomorphism from $L^2(G)$ onto $L^2_\mr{pol}(L\otimes\delta)$.
 \relax
 \bigskip

{\it Quantization of polarized classical observables:}~ A
classical observable $f\in C^\infty(\ctg G)$ is polarized if the
Hamiltonian vector field $X_f$ associated with $f$ satisfies
$[X_f,\Gamma(D)]\subseteq\Gamma(D)$. The operator $\hat f$
associated with $f$ is then defined by
 \beq\label{G-qobs-vert0}
\hat f(\vp \otimes \nu)
 =
 \big(
(\mr i \hbar X_f + \theta(X_f) + f)\vp
 \big)
\otimes \nu
 +
\vp\otimes(\mr i \hbar \mc L_{X_f} \nu)
 \,,~~~~~~
\vp\otimes\nu \in\Gamma_\mr{pol}(L\otimes\delta)\,.
 \eeq
Here, $\mc L_X$ denotes the Lie derivative w.r.t.\ the vector
field $X$ on $\ctg G$, which is defined on sections of $\delta$ by
virtue of the Leibniz rule
 \beq\label{G-defLie}
 \textstyle
(\mc L_{X} \nu) \otimes\nu := \frac12 \mc L_{X} (\nu\otimes\nu)\,.
 \eeq
The first term in \eqref{G-qobs-vert0} contains the ordinary
quantization formula of Kostant and Souriau, whereas the second
term represents the half-form correction. If $X_f$ is complete,
$\hat f$ is essentially self-adjoint \cite{Woodhouse}. The
argument is as follows. For any polarized $f$, the flow of $X_f$
lifts to a flow on $L\otimes\delta$, where the lift to $\delta$ is
natural and the lift to $L$ is defined by the connection $\nabla$.
If $X_f$ is complete, the lifted flow induces a strongly continuous
1-parameter group of unitary transformations on
$L^2_\mr{pol}(L\otimes\delta)$. The self-adjoint generator of this
group, which exists due to Stone's theorem, has the subspace
$\Gamma_\mr{pol}(L\otimes\delta)$ as a core and on this
core it is given by \cite{Woodhouse}:
$$
\hat f(\vp \otimes \nu)
 =
 \big(
(\mr i \hbar X_f + \theta(X_f))\vp
 \big)
\otimes \nu
 +
\vp\otimes(\mr i \hbar \mc L_{X_f} \nu)
 \,,~~~~~~
\vp\otimes\nu \in\Gamma_\mr{pol}(L\otimes\delta)\,.
$$
By adding the multiplication operator  by the real function $f$ we obtain  
$\hat f$ as an essentially self-adjoint operator. 

By virtue of the isomorphism between $L^2(G)$ and
$L^2_\mr{pol}(L\otimes\delta)$ defined by \eqref{G-ismSrqiz},
$\hat f$ is mapped to an essentially self-adjoint operator on 
$L^2(G)$ which will be denoted by $\hat f$ as well. According to 
\eqref{G-qobs-vert0}, the defining equation for this operator is
 \beq\label{G-qobs-vert}
\pi^\ast(\hat f \psi) \otimes \sqrt{\pi^\ast v}
 =
 \big(
(\mr i \hbar X_f + \theta(X_f) + f) \pi^\ast\psi
 \big)
 \otimes\sqrt{\pi^\ast v}
 +
\pi^\ast\psi\otimes(\mr i \hbar \mc L_{X_f} \sqrt{\pi^\ast v})\,,
 \eeq
where $\psi\in C^\infty(G)$.
 \bigskip

{\it Quantization of $f_0$ and $f_1$:}~ We check that $f_0$ and $f_1$ are
polarized. Since in the decomposition \eqref{G-decotg}, elements
of $\Gamma(D)$ are characterized by having zero first component,
it suffices to take the commutator of $X_{f_i}$ with the constant
vector fields $(0,B)$ where $B\in\mf g$. Since, according to
\eqref{G-Havf0} and \eqref{G-Havf1}, the first components of
$X_{f_0}$ and $X_{f_1}$ do not depend on the momentum variable
$X$, only their second components can contribute to the commutator
$[X_{f_i},(0,B)]$. Since $\Gamma(D)$ is integrable, then
$[X_{f_i},(0,B)]\in\Gamma(D)$, $i=0,1$.

Next, we determine $\hat f_0$ and $\hat f_1$ using \eqref{G-qobs-vert}.
For $f = f_0$, \eqref{G-splpot} and \eqref{G-Havf0}
yield $\theta(X_{f_0}) = 0$ as well as $X_{f_0} \pi^\ast \psi= 0$ and $\mc
L_{X_{f_0}} \pi^\ast v = 0$, hence $\mc L_{X_{f_0}}\sqrt{\pi^\ast v} = 0$. Thus,
\eqref{G-qobs-vert} yields
 \beq\label{G-qobs0}
 \textstyle
\hat f_0 \psi = f_0\psi
 \,,~~~~~~
\psi\in C^\infty(G)\,,
 \eeq
where on the r.h.s., $f_0$ is viewed as a function on $G$ rather than on $\ctg
G$. As $G$ is compact, $f_0$ is bounded, hence $\hat f_0$ extends to a bounded
self-adjoint operator on $L^2(G)$, which will be denoted by the same symbol.

For $f = f_1$, \eqref{G-splpot} and \eqref{G-Havf1} yield
$
\theta_{(a,X)}(X_{f_1})
 =
\langle X , \OPg(a) \rangle
 =
- \frac{1}{2\scale^2} \tr(Xa)
 =
- f_1(a,X)\,.
$
Hence,
 \beq\label{G-f1-1}
\theta\big(X_{f_1}\big) + f_1 = 0\,.
 \eeq
Furthermore, we observed before that the first component of $X_{f_1}$ in
the decomposition \eqref{G-decotg} does not depend on the momentum variable $X$.
Hence, this component defines a vector field $Y_{f_1}$ on $G$. According to
\eqref{G-Havf1},
 \beq\label{G-Yi}
\big(Y_{f_1}\big)_a = \mr L_a' \OPg(a)
 \,,~~~~~~
a\in G\,.
 \eeq
By construction,
 \begin{align} \label{G-f1-2}
X_{f_1}\pi^\ast\psi & = \pi^\ast\big(Y_{f_1}\psi\big)\,,
\\ \label{G-f1-3}
\mc L_{X_{f_1}}\pi^\ast v & = \pi^\ast \big(\mc L_{Y_{f_1}} v\big)\,.
 \end{align}
A straightforward computation yields
 \beq\label{G-f1-4}
\mc L_{Y_{f_1}} v = \frac{3}{2} f_0 \, v\,,
 \eeq
see the appendix. Then \eqref{G-defLie} and \eqref{G-f1-3} yield
$\mc L_{X_{f_1}} \sqrt{\pi^\ast v} = \frac{3}{4} f_0 \, \sqrt{\pi^\ast v}$.
Plugging in this as well as \eqref{G-f1-1} and \eqref{G-f1-2} into
\eqref{G-qobs-vert} we arrive at
 \beq\label{G-qobs1}
 \textstyle
\hat f_1\psi = \mr i \hbar \left(Y_{f_1} + \frac34 \hat f_0\right)\psi
 \,,~~~~~~\psi\in C^\infty(G)\,.
 \eeq
Since according to Lemma \rref{L-Havf}, the Hamiltonian vector field $X_{f_1}$ 
is complete, $\hat f_1$ is essentially self-adjoint. From now on, $\hat f_1$
will denote the self-adjoint extension. Thus, all the operators $\hat f_0$,
$\hat f_1$ and $\hat f_2$ are self-adjoint and have $C^\infty(G)$ as a common
invariant core. 
\bre

Consider the operator $\mr i\hbar Y_{f_1}$ on $C^\infty(G)$. Since $\hat f_1$
and $\hat f_0$ are symmetric, $\mr i\hbar Y_{f_1} = \hat f_1  - \mr
i\hbar\frac34\hat f_0$ is not. Thus, the term $\mr i\hbar \frac34 \hat f_0$,
playing the role of the half-form correction in the quantization of $f_1$, can
be characterized as the unique purely imaginary multiplication operator which
has to be added to the 'na\"ive quantization' $\mr i\hbar Y_{f_1}$ of $f_1$ in
order to obtain a symmetric operator.

\ere

Finally, by reduction after quantization we arrive at the
Hilbert space $L^2(G)^G$ of $G$-invariant elements. Since the functions
$f_i$ are $G$-invariant, by restriction, the operators $\hat f_i$
define self-adjoint operators on $L^2(G)^G$ which will be denoted by
the same symbols. The subspace $C^\infty(G)^G$ is a common invariant core for
these operators.

An orthonormal basis in $L^2(G)^G$ is
provided by the real characters $\chi_n$, where $n=0,1,2,\dots$ is twice the
spin and labels the irreducible representations of $G$. To have the
formulae in the following proposition valid for all $n$, let $\chi_{-1} = 0$.

 \bpr\label{P-qobs-bs}

In the basis of characters, the quantum observables $\hat f_i$ are
given by
 \begin{eqnarray}\label{G-qobs-bs-0}
 \textstyle
\hat f_0 \chi_n
 & = &
 \textstyle
\chi_{n+1} + \chi_{n-1}\,,
\\ \label{G-qobs-bs-1}
 \textstyle
\hat f_1 \chi_n
 & = &
 \textstyle
\mr i \hbar \left(\frac{2n+3}{4} \chi_{n+1}
- \frac{2n+1}{4}\chi_{n-1}\right)\,,
 \textstyle
\\ \label{G-qobs-bs-2}
\hat f_2 \chi_n
 & = &
 \textstyle
\hbar^2 \scale^2 n(n+2) \chi_n\,.
 \end{eqnarray}
Accordingly, their matrix elements are
 \begin{eqnarray*}
(\hat f_0)_{nm}
 & = &
 \textstyle
\delta_{n\,m+1} + \delta_{n\,m-1}\,,
\\
(\hat f_1)_{nm}
 & = &
 \textstyle
\mr i\hbar \left(\frac{2m+3}{4}
\delta_{n\,m+1} - \frac{2m+1}{4} \delta_{n\,m-1}\right)\,,
\\
(\hat f_2)_{nm}
 & = &
 \textstyle
\hbar^2 \beta^2 \frac{n(n+2)}{2} \delta_{nm}\,.
 \end{eqnarray*}
 \verweis{(SQMG-244)}%

\epr

{\it Proof.}~ As $\hat f_2$ is the negative of the Laplacian on
$G$, the formula for $\hat f_2$ is standard. As $\hat f_0$ is
multiplication by $f_0$ and $f_0$ is the character of the
fundamental representation, the formula for $\hat f_0$ reflects
the ordinary reduction formula for tensor products. For $\hat
f_1$, it suffices to determine $(\hat f_1\chi_n)(a)$ for $a\in T$.
Write $a = \diag(\alpha,\ol\alpha)$ with $\alpha\in\mr U(1)$. Then
$$
\chi_n(a)
 =
\alpha^n + \alpha^{n-2} + \cdots + \alpha^{-n}\,.
$$
We compute
 $
 \textstyle
\left(Y_{f_1}\right)_a \chi_n
 =
\ddtn \chi_n\left( a \mr e^{\OPg(a)t} \right)
 $.
Since
 $
a \mr e^{\OPg(a)t} = \diag\left(\alpha \mr e^{\frac12(\alpha
- \ol\alpha)t} , \ol\alpha \mr e^{-\frac12(\alpha -
\ol\alpha)t}\right)
 $,
we have
$$
\chi_n\left( a \mr e^{\OPg(a)t} \right)
 =
\alpha^n \mr e^{\frac n 2(\alpha - \ol\alpha)t}
 +
\alpha^{n-2} \mr e^{\frac {n-2} 2(\alpha - \ol\alpha)t}
 + \cdots +
\alpha^{-n} \mr e^{-\frac n 2 (\alpha - \ol\alpha)t}\,.
$$
Taking the derivative and sorting by powers of $\alpha$ we obtain
$$
 \textstyle
\left(Y_{f_1}\right)_a \chi_n
 =
\frac n 2 \alpha^{n+1} - \alpha^{n-1} - \alpha^{n-3}
 - \cdots -
\alpha^{-n+1} + \frac n 2 \alpha^{-n-1}
 =
\frac n 2 \chi_{n+1}(a) - \frac {n+2} 2 \chi_{n-1}(a)\,.
$$
Combining this with \eqref{G-qobs-bs-0} we arrive at \eqref{G-qobs-bs-1}.
 \qed
 \bigskip

\bre

Composition of the trivialization $\ctg G\cong G\times \mf g$ with
the inverse of the polar decomposition on $G^\CC$ yields a natural
diffeomorphism $\ctg G\cong G^\CC$. In our situation, $G^\CC = \SL(2,\CC)$. By
virtue of this diffeomorphism, the complex structure of $G^\CC$ and the
symplectic structure of $\ctg G$ combine to a K\"ahler structure. Therefore, in
addition to the vertical polarization defined by the fibres, $\ctg G$ carries a
canonical K\"ahler polarization defined by the K\"ahler structure. For
quantization in this polarization (K\"ahler quantization) on a general compact
Lie group see \cite{Hall:Compact}. The Hilbert space $\mc H_{\text{K\"ahler}}$
of K\"ahler quantization consists of holomorphic function on $G^\CC$ which are
square-integrable w.r.t.\ a certain measure. Since the elements of $\mc
H_{\text{K\"ahler}}$ are true functions rather than classes of functions, it is
this space on which one constructs the costratied Hilbert space structure that
implements the stratification of the reduced phase space on the level of the
quantum theory, see \cite{Hue:qr} and \cite{hrs}.
There exists a natural unitary isomorphism between the Hilbert spaces of the
Schr\"odinger and the K\"ahler quantization ('generalized Bargmann-Segal
transformation'). For general compact $G$, this isomorphism was first given in
\cite{Hall:Compact} in terms of the heat kernel on $G^\CC$. Later on, in
\cite{Hue:holopewe} a Peter-Weyl theorem for the Hilbert space of K\"ahler
quantization was proved and it was used to show that the isomorphism between the
two Hilbert spaces can also be obtained by matching irreducible components of
the standard $G\times G$-representations on these two Hilbert spaces. For the
subspaces of invariants this implies that here the unitary isomorphism is given
by mapping each $\chi_n$ to the corresponding character on $G^\CC$, normalized
w.r.t.\ the specific scalar product on $\mc H_{\text{K\"ahler}}$.

\ere


\subsection{Domains, eigenvalues and spectra of the quantized generators}


To investigate the operators $\hat f_i$ we pass from $L^2(G)^G$ to $L^2[0,\pi]$
as follows. Let $C^\infty[0,\pi]$ denote the Whitney
smooth functions on the closed interval $[0,\pi]$ (i.e., smooth functions on the
open interval $]0,\pi[$ that can be smoothly extended outside $[0,\pi]$). Take
the parameterization $\phi$ of the subgroup $T\subseteq G$ of diagonal matrices,
see \eqref{G-paramT}, and define a map $\Gamma : C^\infty(G) \to
C^\infty[0,\pi]$ by
$$
(\Gamma \psi)(x) = \sqrt2 ~ \sinfn x ~ \psi\big(\phi(x)\big)
 \,,~~~~~~
x\in[0,\pi]\,.
$$

\ble\label{L-L2}

$\Gamma$ extends to a unitary Hilbert space isomorphism $L^2(G)^G\to
L^2[0,\pi]$.

\ele

{\it Proof.}~ We have to check that $\Gamma$ is isometric and that its image is
dense in $L^2[0,\pi]$. Let $\psi,\vp\in L^2(G)^G$. From the Weyl integration
formula we know that
$$
 \textstyle
\int_G \overline\psi \vp \mr da
 =
\int_T \overline\psi \vp v\mr dt\,,
$$
where $\mr da$ and $\mr dt$ denote the Haar measures on $G$ and $T$,
respectively, and $v$ is a density function that accounts
for the volume of the orbits under inner automorphisms of $G$. For $G = \SU(2)$,
$$
 \textstyle
\phi^\ast(v\mr d t) = \frac{\vol(G)}{\pi} \sinfnpot 2 x\mr dx\,.
$$
Hence,
 \begin{align*}
\braket{\psi}{\vp}
 & =
 \textstyle
\frac{1}{\vol(G)} \int_T \overline{\psi} \, \vp v\, \mr dt
\\
 & =
 \textstyle
\frac{1}{\vol(G)} \int_{-\pi}^\pi \phi^\ast
 \left(\overline{\psi} \, \vp \, v \, \mr dt\right)
\\
 & =
 \textstyle
\frac{1}{\pi} \int_{-\pi}^\pi
 ~
\overline{\psi\big(\phi(x)\big)}
 ~
\vp\big(\phi(x)\big)
 ~
\sinfnpot 2 x \, \mr dx\,.
\\
 & =
 \textstyle
\frac{1}{2\pi} \int_{-\pi}^\pi
 ~
\overline{(\Gamma\psi)(x)}
 ~
(\Gamma\vp)(x) ~ \mr dx\,.
 \end{align*}
Since $\psi$ and $\vp$ are invariant under inner automorphisms, $\Gamma\psi$ and
$\Gamma\vp$ are invariant under reflection $x\mapsto -x$. Hence, the integral
over $[-\pi,\pi]$ gives twice the integral over $[0,\pi]$. This shows that
$\Gamma$ is isometric. Since the image of $\Gamma$ contains the smooth
functions with compact support inside the open interval $]0,\pi[$, it is dense
in $L^2[0,\pi]$.
 \qed
\\

We will need the image of $C^\infty(G)^G$ under $\Gamma$. Let $\cdic$ denote the
subspace of $C^\infty[0,\pi]$ of functions whose even order derivatives
$\psi^{(2n)}$, $n=0,1,2,,\dots$, vanish in $0$ and $\pi$:
$$
\cdic = \{\psi\in C^\infty[0,\pi] : \psi^{(2n)} (0) = \psi^{(2n)}
(\pi) = 0\,,~n=0,1,2,\dots\}\,.
$$

\ble\label{L-core}

$\Gamma\big(C^\infty(G)^G\big) = \cdic$.

\ele

{\it Proof.}~ First, let $\vp\in C^\infty(G)^G$. Define $\tilde\vp\in
C^\infty(\RR)$ by $\tilde\vp(x) := \vp\big(\phi(x)\big)$. Then $\Gamma(\vp)(x) =
\sqrt 2 \sinfn x \tilde\vp(x)$ and the iterated Leibniz rule yields for the
derivative of order $2n$
 \begin{align*}
 \textstyle
\Gamma(\vp)^{(2n)}(x)
 & =
 \textstyle
\sqrt2
 \Big\{
\sum_{k=0}^n (-1)^{n-k} {2n \choose 2k} \sinfn x \tilde\vp^{(2k)}(x)
\\
 & \hspace{5cm} +
 \textstyle
\sum_{k=0}^{n-1} (-1)^{n-k-1} {2n \choose 2k+1} \cosfn x
\tilde\vp^{(2k+1)}(x)\,.
 \Big\}
 \end{align*}
By construction, the function $\tilde\vp$ is $2\pi$-periodic and has even
parity, i.e., $\tilde\vp(-x) = \tilde\vp(x)$. Hence, the derivative $\vp^{(k)}$
is $2\pi$-periodic and has even parity for even $k$ and odd parity for odd $k$.
It follows $\vp^{(2n+1)} (0) = \vp^{(2n+1)} (\pi) = 0$ and hence
$\Gamma(\vp)^{(2n)}(0) = \Gamma(\vp)^{(2n)}(\pi) = 0$, for any $n$.

Conversely, let $\psi\in\cdic$. Since $\psi(0) = \psi(\pi) = 0$, we can extend
$\psi$ to a well-defined function on the whole of $\RR$ by setting $\psi(-x) =
-\psi(x)$ and $\psi(x+2\pi m) = \psi(x)$, $x\in[0,\pi]$, $m$ an integer. Then
for any $x\in\RR\setminus2\pi\ZZ$, any $k=0,1,2,\dots$ and any $m\in\ZZ$ there
holds $\psi^{(k)}(-x) = -(-1)^k \psi^{(k)}(x)$ and $\psi^{(k)}(x+2\pi m) =
\psi^{(k)}(x)$. In addition, $\lim_{x\to2\pi m} \psi^{(2k)}(x) = 0$. This
implies that derivatives of $\psi$ of arbitrary order are continuous in $x =
2\pi m$, hence $\psi$ is smooth. Now define a function $\tilde\vp$ on
$\RR\setminus2\pi\ZZ$ by $\tilde\vp(x) = \frac{1}{\sqrt
2}\,\frac{\psi(x)}{\sinfn x}$. We
claim that $\tilde\vp$ extends to a smooth function on the whole of $\RR$. To
see this, it suffices to show smoothness in $x=0$ and $x=\pi$. We give the
argument for $x=0$ only; the case $x=\pi$ is analogous. Since the sine function
is a local diffeomorphism in a neighbourhood of $x=0$, $\tilde\vp$ is smooth in
$0$ iff so is $\tilde\vp\circ\arcsin$. Denote $f(x) = \psi\big(\arcsinfn
x\big)$. Then $f$ is smooth in a neighbourhood of $x=0$ and
$\tilde\vp\circ\arcsin(x) = \frac{1}{\sqrt 2}\,\frac{f(x)}{x}$. Hence,
$$
 \textstyle
(\tilde\vp\circ\arcsin)^{(k)}(x)
 =
\frac{1}{\sqrt 2}\,\frac{1}{x^{n+1}}
 ~
\sum_{l=0}^k ~ (-1)^{n-k} \, {k \choose l} \, (k-l)! \, x^k \, f^{(k)}(x)\,.
$$
Since $f(0) = 0$, the r.h.s.\ yields an indefinite expression for $x\to 0$.
The derivative of the enumerator is $x^k f^{(k+1)}(x)$. Hence, the rule
of de l'Hospital yields
$$
 \textstyle
\lim_{x\to 0}(\tilde\vp\circ\arcsin)^{(k)}(x)
 =
\frac{1}{\sqrt2} \, \frac{f^{(k+1)}(0)}{k+1}\,.
$$
This proves that $\tilde\vp$ extends to a smooth function on $\RR$. Since it is
$2\pi$-periodic and has even parity by construction, there exists $\vp\in
C^\infty(G)^G$ such that $\tilde\vp = \vp\circ\phi$. Then $\psi = \Gamma(\vp)$.
 \qed
 \\

The operators $\hat f_i$ on $L^2(G)^G$ induce operators $\Gamma \hat f_i
\Gamma^{-1}$ on $L^2[0,\pi]$. These induced operators will also be denoted by
$\hat f_i$. We derive explicit expressions. Since $\Gamma \hat f_0 \Gamma^{-1}$
is multiplication by the function $f_0\circ\phi$,
 \beq\label{G-f0expr}
\hat f_0\psi(x) = 2\cosfn{x} \, \psi(x)\,.
 \eeq
Let $\AC[0,\pi]$ denote the space of absolutely
continuous functions and let
 \begin{align*}
\AC^1[0,\pi] & = \{\psi\in\AC[0,\pi] : \psi'\in L^2[0,\pi]\}\,,
 \\
\AC^2[0,\pi] & = \{\psi\in\AC^1[0,\pi] : \psi'\in\AC^1[0,\pi]\}\,.
 \end{align*}

\bpr\label{P-dom}~

The operator $\hat f_1$ has domain
$
\mr D(\hat f_1)
 =
\{\psi\in L^2[0,\pi] : \sinfn{x}\psi(x) \in \AC^1[0,\pi]\}
$
and is given by the expression
 \beq\label{G-f1expr}
 \textstyle
\hat f_1 = \mr i \hbar \left(\ddx \sinfn{x} \,-\, \frac12\cosfn x\right)\,.
 \eeq
The operator $\hat f_2$ has domain
$
\mr D(\hat f_2)
 =
\{\psi\in L^2[0,\pi] : \psi \in \AC^2[0,\pi], \psi(0) = \psi(\pi) = 0\}
$
and is given by the expression
 \beq\label{G-f2expr}
 \textstyle
\hat f_2 = - \hbar^2\beta^2\left(\ddxx + 1\right)\,.
 \eeq
The subspace $\cdic$ is a common invariant core for $\hat f_0$, $\hat
f_1$, $\hat f_2$.

\epr

\bre

For $\psi\in C^\infty[0,\pi]$, one has
 \begin{align}\nonumber
 \textstyle
\hat f_1 \psi
 & =
 \textstyle
\mr i \hbar \left(\ddx \sinfn{x} \,-\, \frac12\cosfn x\right) \psi(x)
\\ \nonumber
 & =
 \textstyle
\mr i \hbar \left(\sqrt{\sinfn{x}}\ddx\sqrt{\sinfn{x}}\right) \psi(x)
\\ \label{G-f1variants}
 & =
 \textstyle
\mr i \hbar \left(\sinfn{x}\ddx \,+\, \frac12\cosfn x\right) \psi(x)\,,
 \end{align}
whereas it is only the first of these three expressions that extends to the
whole of $\mr D(\hat f_1)$.

 \ere

 \bre

For general $\SU(n)$ the Hilbert space $L^2(G)^G$ can be realized as
$L^2(\sigma^{n-1},v\rm dt)$, where $\sigma^{n-1}$ is the $(n-1)$-simplex (more
concretely, a Weyl alcove in the Lie algebra $\mf g$) and $v\mr dt$ is an
appropriate measure on $\sigma^{n-1}$. In \cite{Wren} it is proved
that in this  realization a core for the group Laplacian $\Delta_G$
is given by Neumann boundary conditions at the boundary of $\sigma^{n-1}$. In
our situation, $L^2(\sigma^{n-1},v\mr dt)$ corresponds to $L^2([0,\pi],\sinfnpot
2 x \mr dx)$ and the core isolated in \cite{Wren} amounts to $\{\psi\in C^\infty[0,\pi]
: \psi'(0) = \psi'(\pi) = 0\}$. By means of the isomorphism $\Gamma$, this core
is mapped into $\{\psi\in C^\infty[0,\pi] : \psi(0) = \psi(\pi) = 0\}$. Thus, the
assertion about the core for $\hat f_2$ in Proposition \rref{P-dom} is
consistent with the result of \cite{Wren}.

 \ere

 \bre

The domain of $\hat f_1$ is unusually large for a differential operator,
it contains e.g.\ all smooth functions. This is due to the fact that, in $\hat
f_1$, the derivative is combined with the sine function which destroys any
information about the boundary values of the function whose derivative is taken.
In particular, $C^\infty[0,\pi]$ may also be taken as a core for $\hat f_1$.

 \ere

 \bre

Occasionally we will have to deal with the operator $\hat f_1^2$
below. For further use we note that the domain of $\hat f_1^2$
contains $\AC^2[0,\pi]$ as a proper subspace and that on
$\AC^2[0,\pi]$,
 \beq\label{G-f12expr}
 \textstyle
\hat f_1^2
 =
-\hbar^2
 \left(
\sinfn{x}\ddxx\sinfn{x} - \frac14\cosfnpot 2 x + \frac12
 \right)\,.
 \eeq

 \ere

{\it Proof.}~ The last statement follows from the fact that $C^\infty(G)^G$ is a
common invariant core for $\hat f_0$, $\hat f_1$, $\hat f_2$ and Lemma
\rref{L-core}.

First, consider $\hat f_2$. According to the general formula for the radial part
of the Laplacian on a compact group, see \cite[\S II.3.4]{Helgason}, the
restriction of $\hat f_2$ to $\cdic$ is given by the r.h.s.\ of
\eqref{G-f2expr}. The assertion about the domain then follows by standard
extension theory for the operator of second derivative. Since the r.h.s.\ of \eqref{G-f2expr} is well defined on
$\AC^2[0,\pi]$, $\hat f_2$ is given by this expression on the whole of its
domain.

Next, consider $\hat f_1$. According to \eqref{G-qobs1}, for $\psi\in \cdic$,
$$
 \textstyle
\hat f_1\psi
 \equiv
\Gamma \hat f_1 \Gamma^{-1}\psi
 =
\mr i \hbar \Gamma \big(Y_{f_1} + \frac34 \hat f_0\big) \Gamma^{-1}\psi\,.
$$
According to \eqref{G-Yi} and \eqref{G-defP},
$$
 \textstyle
\big(Y_{f_1}\psi\big)(x)
 =
\sqrt 2 \sinfn x \, \ddtn (\Gamma^{-1}\psi)
 \left(
\phi(x) \mr e^{\frac12\left(\phi(x) - \phi(x)^\dagger\right)t}
 \right)\,.
$$
A brief computation shows
$
\phi(x) \mr e^{\frac12\left(\phi(x) - \phi(x)^\dagger\right)t}
 =
\phi(x + t \sinfn x)\,.
$
Hence,
$$
 \textstyle
\big(Y_{f_1}\psi\big)(x)
 =
\sqrt 2 \sinfn x \, \ddtn \frac{\psi(x + t \sinfn x)}{\sqrt2\sin(x + t \sinfn x)}
 =
\psi'(x) \sinfn x - \psi(x) \cosfn x\,.
$$
Together with \eqref{G-f0expr} and \eqref{G-f1variants} this yields
$
\big(\hat f_1\psi\big)(x)
 =
\mr i \hbar \left(\ddx \sinfn{x} \,-\, \frac12\cosfn x\right) \psi(x)\,,
$
hence on $\cdic$, $\hat f_1$ is given by the r.h.s.\ of \eqref{G-f1expr}. Denote $D = \{\psi\in L^2[0,\pi] : \sinfn{x}\psi(x)
\in \AC^1[0,\pi]\}$. Since the r.h.s.\ of \eqref{G-f1expr} is well-defined for
all $\psi\in D$, $\hat f_1$ is given by this expression on the whole of $D$. It
remains to show $\mr D(\hat f_1) = D$.

Let $A$ be defined by restriction of $\hat f_1$ to the core $\cdic$. Since $f_1$ is self-adjoint, 
$$
A^\dagger = \ol A{}^\dagger = \hat f_1^\dagger = \hat f_1 \, .
$$
Hence, it suffices to
show $\mr D(A^\dagger) = D$. Let $\psi\in D$. Then $\sinfn{x}\psi(x)\in
\AC^1[0,\pi]$, hence it has a derivative $(\sinfn{x}\psi(x))'\in L^1[0,\pi]$
and $(\sinfn{x}\psi(x))'\in L^2[0,\pi]$ . Then
$$
 \textstyle
\tilde\psi(x)
 :=
\mr i\hbar \big((\sinfn{x}\psi(x))' - \frac12\cosfn x\psi(x)\big) \in
L^2[0,\pi]\,.
$$
For any $\vp\in \cdic$, integration by parts yields
 \begin{align*}
 \textstyle
\braket{\tilde\psi}{\vp}
 & =
 \textstyle
-\frac{\mr i \hbar}{\pi} \int_0^\pi
 \ol{
\big((\sinfn{x}\psi)' - \frac12\cosfn x\psi(x)\big)
 }
\vp(x) \, \mr dx
\\
 & =
 \textstyle
\frac{\mr i \hbar}{\pi}
 \int_0^\pi
\ol{\psi(x)}\big(\sinfn{x}\vp' + \frac12\cosfn x\vp(x)\big)
  \, \mr dx
\\
 & =
 \textstyle
\braket{\psi}{A\vp}\,,
 \end{align*}
hence $\psi\in\mr D(A^\dagger)$. Conversely, let $\psi\in\mr D(A^\dagger)$. Then
there exists $\tilde\psi\in L^2[0,\pi]$ such that $\braket{\psi}{A\vp} =
\braket{\tilde\psi}{\vp}$ for all $\vp\in \cdic$. Write this equation in the
form
$$
 \textstyle
\int_0^\pi \ol{\sinfn{x}\psi(x)}
 ~
\mr i\ddx \vp(x)
 ~
\mr dx
 =
\int_0^\pi
 \ol{\left(
\frac1\hbar \tilde\psi
 +
\frac{\mr i}{2} \cosfn x\psi(x)
 \right)}
 ~
\vp(x)
 ~
\mr dx
 \,,~~~~~~
\forall~~\vp\in \cdic\,.
$$
We conclude that $\sinfn{x}\psi(x)$ belongs to the domain of the adjoint of the restriction of $\mr i\ddx$ to the 
subspace $\cdic \,. $ Since  $\cdic $ is a core for $\mr i\ddx$ and since the domain of the self-adjoint operator $\mr i\ddx$ is $\AC^1[0,\pi]$ it follows that  $\sinfn{x}\psi(x)\in\AC^1[0,\pi]$, i.e., $\psi\in D \, .$ 
\qed
 \\

Next, we discuss the eigenvalues and the spectra of the operators $\hat f_i$.
According to \eqref{G-qobs-bs-2}, $\hat f_2$ has pure point spectrum,
$$
\sigma(\hat f_2) = \{\hbar^2n(n+2) : n=0,1,2,\dots\}
$$
and the characters form an orthonormal basis of eigenvectors.

\bpr\label{P-spec}

The operators $\hat f_0$, $\hat f_1$ and $\hat f_1^2$ do not possess
eigenvalues. Their spectra are
$$
\sigma(\hat f_0) = [-2,2]
 \,,~~~~~~
\sigma(\hat f_1) = \RR
 \,,~~~~~~
\sigma(\hat f_1^2) = [0,\infty[
 \,.
$$

\epr

{\it Proof.}~ First, consider $\hat f_0$. The eigenvalue equation $(\hat f_0
-\lambda)\psi = 0$ reads $(2\cosfn x-\lambda)\psi(x) = 0$, hence $\psi=0$ a.e.\ for any
$\lambda\in\RR$. Thus, there are no eigenvalues. The assertion about the
spectrum follows from the spectral mapping theorem.

Next, consider $\hat f_1$. According to \eqref{G-f1expr}, the eigenvalue
equation amounts to the differential equation
 \beq\label{G-evaleqf1}
 \textstyle
\big(\hat f_1 - \lambda\big) \psi(x)
 =
\big\{\mr i\hbar\big(\ddx\sinfn x - \frac12 \cosfn x\big) - \lambda\big\} \psi(x)
 =
0
 \eeq
which on the open interval $]0,\pi[$ can be written in the form
$$
 \textstyle
\mr i \hbar
 \left\{
\ddx
 +
\left(\frac{\mr i\lambda}{\hbar\sinfn x} - \frac12\cot(x)\right)
 \right\}
 ~
(\sinfn{x}\psi(x))
 =
0\,.
$$
For any $\lambda\in\RR$ the
solution is
 \beq\label{G-psil}
 \textstyle
\psi_\lambda(x)
 =
\frac{1}{\sqrt{2\hbar}}
 ~
\frac{\mr e^{-\frac{\mr i}{\hbar} \lambda \ln \tanfn{\frac x2}}
 }{
\sqrt{\sinfn x}
 }\,.
\eeq
The particular choice of normalization will be justified below.
Since neither of the functions $\psi_\lambda$ is square
integrable, $\hat f_1$ does not have eigenvalues.

To determine the spectrum of $\hat f_1$, let $\lambda\in\RR$. If $\hat f_1-\lambda$ had a
bounded inverse, there would exist $C>0$ such that $\|\psi\| = \|(\hat
f_1-\lambda)^{-1}(\hat f_1-\lambda)\psi\|
\leq C \|(\hat f_1-\lambda)\psi\|$ for any $\psi\in\mr D(\hat f_1)$. Thus, in
order to show that $\lambda\in\sigma(\hat f_1)$ it
suffices to construct a sequence $\psi_n$ in $\mr D(\hat f_1)$ such that
$\frac{\|\psi_n\|}{\|(\hat f_1-\lambda)\psi_n\|} \to \infty$.
Choose a smooth function $j$ on $\RR$ with support in the open interval
$]-1,1[$ such that $0\leq j(x)\leq 1$ and
$\int_{-\infty}^\infty j(x) \, \mr d x
= 1$. Define $g_n(x) = n\int_{-\infty}^x \big\{ j(nx'-2) - j(n(x'-\pi)+2)
\big\}\,\mr d x'$ and $\psi_n := g_n\psi_\lambda$. Since $g_n$ has
support in $]0,\pi[$, $\psi_n\in L^2[0,\pi]$ and hence $\psi_n\in\mr D(\hat
f_1)$. On the open interval $]0,\pi[$ we have
$$
 \textstyle
\big((\hat f_1-\lambda)\psi_n\big)(x)
 =
\mr i\hbar \big(\ddx g_n\big)(x) \sinfn{x}\psi_\lambda(x)
 +
g_n(x) \left(\mr i\hbar \big(\ddx\sinfn{x} - \frac12\cosfn x\big)-\lambda\right)
\psi_\lambda(x)\,.
$$
The second term vanishes because $\psi_\lambda$ solves \eqref{G-evaleqf1} on
$]0,\pi[$. Hence
$$
 \textstyle
\big((\hat f_1-\lambda)\psi_n\big)(x)
 =
\mr i \sqrt{\frac{\hbar}{2}}
 ~
\big(nj(nx-2) - nj(n(x-\pi)+2)\big)
 ~
\sqrt{\sinfn x}
 ~
\mr e^{-\frac{\mr i}{\hbar}\lambda \ln\tanfn{\frac x2}}
$$
and therefore
$$
 \textstyle
\|(\hat f_1-\lambda)\psi_n\|^2
 =
\frac{\hbar}{2\pi} n^2
 \left\{
\int_0^\pi j(nx-2)^2 \, \sinfn x \, \mr d x
 +
\int_0^\pi j\big(n(x-\pi)+2\big)^2 \, \sinfn x \, \mr d x
 \right\}\,.
$$
The mixed term vanishes for $n$ large enough because $j(nx-2)$ has support
in $]\frac1n,\frac3n[$ and $j(n(x-\pi)+2)$ has support in
$]\pi-\frac3n,\pi-\frac1n[$. For the same reason,
$j(nx-2)\sinfn x\leq\frac3n$ and $j(n(x-\pi)+2)\sinfn x\leq\frac3n$. Hence
$$
 \textstyle
\|(\hat f_1-\lambda)\psi_n\|^2
 \leq
\frac{3\hbar}{2\pi} n
 \left\{
\int_0^\pi j(nx-2) \, \mr d x
 +
\int_0^\pi j\big(n(x-\pi)+2\big) \, \mr d x
 \right\}
 =
\frac{3\hbar}{\pi}\,.
$$
It follows
$$
 \textstyle
\frac{\|\psi_n\|^2}{\|(\hat f_1-\lambda)\psi_n\|^2}
 \geq
\frac{\pi}{3\hbar} ~ \|\psi_n\|^2
 =
\frac{1}{6\hbar^2} \int_0^\pi \frac{g_n^2(x)}{\sinfn x} \, \mr d x
 \to
\infty
$$
and hence $\lambda\in\sigma(\hat f_1)$.

Finally, consider $\hat f_1^2$. According to \eqref{G-f12expr},
the eigenvalue equation $\big(\hat f_1^2 - \lambda^2\big) \psi =
0$ can be written in the form
$$
 \textstyle
-\hbar^2\sinfn x
 \left(
\ddxx
 -
\frac14\cot^2(x)
 +
\left(\frac12 + \frac{\lambda^2}{\hbar^2}\right) \,
\frac{1}{\sinfnpot 2 x}
 \right) ~ (\sinfn{x}\psi(x))
  =
0\,,~~~~~~ x\in]0,\pi[\,,
$$
where it is manifest that for any $\lambda \geq 0$ the solution
space has dimension $2$. Hence, in case $\lambda^2\neq 0$, any
solution is a linear combination of $\psi_\lambda$ and
$\psi_{-\lambda}$ and, therefore, is not square-integrable. Thus,
$\lambda^2$ is not an eigenvalue. In case $\lambda^2 =0$ we
observe that, in addition to $\psi_0(x) =
\frac{1}{\sqrt{\sinfn x}}$, a further solution is given by
$\tilde\psi_0(x) = \frac{\ln\tanfn{\frac x2}}{\sqrt{\sinfn x}}$. Since
neither $\psi_0$ nor $\tilde\psi_0$ is square-integrable,
$\lambda^2 = 0$ is not an eigenvalue, too.

To prove the assertion about the spectrum we choose $\lambda\geq0$
and consider the sequence $\psi_n$ defined above. Obviously,
$\psi_n\in\mr D(\hat f_1^2)$ for all $n$. On $]0,\pi[$ we find, using
\eqref{G-f1expr},
 \begin{align*}
 \textstyle
\big(\hat f_1^2-\lambda^2\big)\psi_n(x)
 & =
 \textstyle
-\hbar^2
 \left\{
g_n'(x) \sinfn{x}
 \left(
\frac12\cosfn{x} - \mr i \frac{\lambda}{\hbar}
 \right)
 +
g_n''(x)\sinfnpot 2 x
 \right\}
\psi_\lambda(x)
\\
 & \hspace{4cm} +
 \textstyle
g_n(x)
 \left(
- \hbar^2
 \left(
\ddx\sinfn{x} - \frac12 \cosfn{x}
 \right)^2
-\lambda^2
 \right)
\psi_\lambda(x)\,.
 \end{align*}
The last term vanishes. Moreover, $g_n''(x) = n^2 \big(j'(nx - 2)
- j'(n(x-\pi)+2)\big)$. Hence,
 \begin{align*}
 \textstyle
\big(\big(\hat f_1^2-\lambda^2\big)\psi_n\big) (x)
 & =
 \textstyle
-\sqrt{\frac{\hbar^3}{2}}
 \left\{
n\big(j(nx-2) - j(n(x-\pi)+2)\big) \sqrt{\sinfn{x}}
 \left(
\frac12\cosfn{x} - \mr i \frac{\lambda}{\hbar}
 \right)\right.
\\
 & \hspace{2cm}+ \left.
 \textstyle
n^2 \big(j'(nx-2)-j'(n(x-\pi)+2)\big) \sinfnpot{\frac32}{x}
 \right\}
\mr e^{-\frac{\mr i}{\hbar} \ln\tanfn{\frac x2}}\,.
 \end{align*}
Consequently, there are two contributions to $\big\|\big(\hat
f_1^2-\lambda^2\big)\psi_n\big\|^2$. One is centered near $x=0$ and is given by
 \beq\label{G-specf12-1}
 \textstyle
\frac{\hbar^3}{2\pi}
 \int_0^\pi
 \Big\{
\frac{\lambda^2}{\hbar^2} n^2 j(nx-2)^2 \sinfn{x}
 +
 \Big(
\frac12 n j(nx-2) \cosfn x + n^2 j'(nx-2) \sinfn{x}
 \Big)^2
\sinfn{x}
 \Big\}
 \,
\mr dx\,,
 \eeq
the other one is centered near $x=\pi$ and is obtained by replacing
$j(nx-2)$ by $j(n(x-\pi)+2)$ and $j'(nx-2)$ by $j'(n(x-\pi)+2)$ in
\eqref{G-specf12-1}. The first term
in \eqref{G-specf12-1} already appeared in the discussion of the
spectrum of $\hat f_1$, hence we know that it is bounded for all
$n$. The integrand of the second term has support in $[\frac1n,\frac3n]$.
There exists $C>0$ such that $|j'(y)|\leq C$ for all $y\in\RR$. Then
$$
 \textstyle
 \big(
\frac12 n j(nx-2) \cosfn x + n^2 j'(nx-2) \sinfn{x}
 \big)^2
\sinfn{x}
 \leq
 \big(
\frac12 n + n^2 C \frac3n
 \big)^2
 \,
\frac3n
 \leq
\left(\frac12+3C\right)^2 n\,,
$$
hence upon integration, this term is bounded by $2\left(\frac12+3C\right)^2$. An
analogous argument applies to the contribution centered near $x=\pi$. Thus,
$\|\big(\hat f_1^2-\lambda^2\big)\psi_n\|$ is bounded for all $n$.
Then $\frac{\|\psi_n\|}{\|(\hat
f_1^2-\lambda^2)\psi_n\|}\to\infty$ and hence $\lambda^2$ belongs
to the spectrum of $\hat f_1^2$.
 \qed
\bigskip

For any $\lambda\in\RR$, the function $\psi_\lambda$ defines a linear functional
on the subspace $C^\infty_0]0,\pi[$ of $L^2[0,\pi]$ by
$$
\braket{\psi_\lambda}{\vp}
 :=
\frac1\pi\int_0^\pi \ol{\psi_\lambda(x)} \, \vp(x) \, \mr d x
 \,,~~~~~~
\vp\in C^\infty_0]0,\pi[\,.
$$
Integration by parts yields $\braket{\psi_\lambda}{(\hat f_1-\lambda)\vp} =
0$ for all $\vp\in C^\infty_0]0,\pi[$, so that $\psi_\lambda$ can be viewed as
a generalized eigenvector of $\hat f_1$.

\bpr\label{P-genevecf1}

The set of generalized eigenvectors $\{\psi_\lambda : \lambda\in\RR\}$ of $\hat
f_1$ is complete and orthogonal in the distributional sense, i.e.,
 \begin{align} \label{G-complete}
\int_{-\infty}^\infty \ol{\psi_\lambda(x)} \, \psi_\lambda(y)
 \,
\mr d\lambda
 & =
\pi \, \delta(x-y)\,,
 &&
x,y\in]0,\pi[\,,
\\ \label{G-ogon}
\frac1\pi\int_0^\pi
\ol{\psi_{\lambda}(x)} \, \psi_{\mu}(x) \, \mr dx
 & =
\delta(\lambda - \mu)\,,
 &&
\lambda,\mu\in\RR\,.
 \end{align}
The assignment of $\tilde\vp(\lambda) =
\braket{\psi_\lambda}{\vp}$ to $\vp\in C^\infty_0]0,\pi[$ extends to a unitary
isomorphism of $L^2[0,\pi]$ onto $L^2(\RR)$ with inverse $\vp(x) =
\int_{-\infty}^\infty \psi_\lambda(x) \, \tilde\vp(\lambda) \, \mr d \lambda$.

\epr

{\it Proof.}~ For $x,y\in]0,\pi[$,
$$
 \textstyle
\int_{-\infty}^\infty \ol{\psi_\lambda(x)}
 \,
\psi_\lambda(y)
 \,
\mr d\lambda
 =
\frac{1}{2\hbar} ~ \int_{-\infty}^\infty  \frac{\mr e^{\frac{\mr
i}{\hbar}\lambda (\ln\tanfn{\frac x2}-\ln\tanfn{\frac y2})}}{\sqrt{\sinfn x\sinfn y}}
 \,
\mr d\lambda
 =
\pi ~
 \frac{
\delta\big(\ln\tanfn{\frac x2}-\ln\tanfn{\frac y2}\big)
 }{
\sqrt{\sinfn x\sinfn y}
 }\,.
$$
The argument of the $\delta$-distribution, viewed as a function of
$x$ with parameter $y$, has derivative $\frac{1}{\sinfn x}$ and a
single zero at $x=y$. Hence, $ \delta\big(\ln\tan{\frac x2}
 -
\ln\tan{\frac y2}\big)
 =
\sinfn{y} \, \delta(x-y)\,.
$
This proves \eqref{G-complete}. For $\lambda,\mu\in\RR$, the substitution $y =
\frac1\hbar \ln\tanfn{\frac x2}$ yields
$$
 \textstyle
\frac1\pi \int_0^\pi \ol{\psi_{\lambda}(x)} \, \psi_{\mu}(x) \, \mr dx
 =
\frac{1}{2\pi\hbar} \int_0^\pi
 \frac{
\mr e^{\frac{\mr i}{\hbar}(\lambda - \mu)\ln\tanfn{\frac x2}}
 }{
\sinfn x
 } \, \mr d x
 =
\frac{1}{2\pi} \int_{-\infty}^\infty
\mr e^{\mr i (\lambda - \mu) y} \, \mr dy
 =
\delta(\lambda-\mu)\,,
$$
hence \eqref{G-ogon}. To prove the last assertion, we observe that
\eqref{G-complete} implies
$
\int_{-\infty}^\infty \ol{\tilde\vp_1(\lambda)} \, \tilde\vp_2(\lambda)
 \,
\mr d\lambda
 =
\braket{\vp_1}{\vp_2}
$
for all $\vp_1,\vp_2\in C^\infty_0]0,\pi[$.
Hence, $\tilde\vp(\lambda)\in L^2(\RR)$ and the assignment $\vp\mapsto\tilde\vp$
extends to an isometric map $L^2[0,\pi]\to L^2(\RR)$. It remains to check that
this map has dense image. Let $\vp\in C^\infty_0(\RR)$. Since the integrals
$\int_{-\infty}^\infty \psi_\lambda(x)\vp(\lambda)\,\mr d\lambda$ exist for any
$x\in]0,\pi[$, they define a function $\vp_0$ on $]0,\pi[$. Due to
\eqref{G-ogon},
$
\frac1\pi \int_0^\pi |\vp_0(x)|^2 \, \mr dx
 =
\int_{-\infty}^\infty |\vp(x)|^2 \, \mr d\lambda\,,
$
hence $\vp_0\in L^2[0,\pi]$. This proves that the extended map is a unitary
isomorphism.
 \qed
 \\


\subsection{Quantum analogue of the classical relation between generators}
\label{S-q-SS-rel}


Consider the relation \eqref{G-rel} and the inequality
\eqref{G-ineq} satisfied by the classical generators $f_i$. Since
$\hat f_2 \geq 0$, the inequality \eqref{G-ineq} has an obvious
quantum counterpart. Concerning the relation \eqref{G-rel}, we
recall that on the domain of $\hat f_2$, $\hat f_1^2$ is given by
\eqref{G-f12expr}. Expressing the r.h.s.\ of this equation in terms of the
$\hat f_i$ we obtain
$$
 \textstyle
\hat f_1^2
 =
\frac{1}{4\beta^2} \sqrt{4-\hat f_0^2} \, \hat f_2 \, \sqrt{4-\hat f_0^2}
 -
\frac{3\hbar^2}{16} \hat f_0^2 + \frac{\hbar^2}{2}\,.
$$
This can be written in the form
 \beq\label{G-Asqrt-f1}
 \textstyle
\sqrt{4-\hat f_0^2} \, \hat f_2 \, \sqrt{4-\hat f_0^2} - 4\beta^2 \hat
f_1^2
 =
\hbar^2\beta^2\left(\frac34\hat f_0^2 - 2\right)\,.
 \eeq
We observe that when replacing $\hat f_i$ by $f_i$,
\eqref{G-Asqrt-f1} reproduces the relation \eqref{G-rel} in the limit $\hbar \to
0$. We will now derive a relation between the quantum observables
$\hat f_0$, $\hat f_1$, $\hat f_2$ which {\em exactly} reproduces
the classical relation \eqref{G-rel} under the na\"ive replacement
of the operators $\hat f_i$ by the phase space functions $f_i$,
$i=0,1,2$. The attribute 'na\"ive' shall remind us that this
operation is well-defined on the level of formal expressions in
the variables $\hat f_0$, $\hat f_1$, $\hat f_2$ but not
necessarily on the level of the operators defined by these expressions.
To begin with, let $A_1$ be the operator defined on $\mr D(\hat f_2)$ by
the l.h.s.\ of
\eqref{G-Asqrt-f1}. Since the domain of $\hat f_2$ is invariant under $\hat f_0$
and contained in the domain of $\hat f_1^2$, on $\mr D(\hat f_2)$ we can define, additionally,  
the following operators:
 \begin{align*}
 \textstyle
A_2 & := 4 \hat f_2- \hat f_0^2 \hat f_2 - 4\beta^2 \hat f_1^2\,,
\\
A_3 & := 4 \hat f_2 - \hat f_0 \hat f_2 \hat f_0 - 4\beta^2 \hat f_1^2\,,
\\
A_4 & := 4 \hat f_2 - \hat f_2 \hat f_0^2 - 4\beta^2 \hat f_1^2\,.
 \end{align*}
Like $A_1$, these operators correspond to the classical phase space function
$(4-f_0^2)f_2 - 4\beta^2 f_1^2$, but contrary to $A_1$ they are polynomial in
the $\hat f_i$. A straightforward computation using
\eqref{G-f0expr}--\eqref{G-f2expr} yields that on $\mr D(\hat f_2)$ there
holds
 \begin{eqnarray}\label{G-A002-f1}
 \textstyle
\frac12(A_2 + A_4)
 & = &
 \textstyle
\hbar^2\scale^2\left(-\frac14 \hat f_0^2 - 2\cdot\II\right)\,,
 \\ \label{G-A020-f1}
 \textstyle
A_3
 & = &
 \textstyle
\hbar^2\scale^2 \left( \frac34 \hat f_0^2 - 6\cdot\II\right)\,,
 \end{eqnarray}
First, we observe that,
similar to \eqref{G-Asqrt-f1}, when replacing $\hat f_i$
by $f_i$, both \eqref{G-A002-f1} and \eqref{G-A020-f1} reproduce the relation
\eqref{G-rel} in the limit $\hbar \to 0$. In addition, we observe that the three
operators $A_1$, $\frac12(A_2 + A_4)$ and $A_3$ are contained in the real
vector space spanned by $\II$ and $\hat f_0^2$. A brief computation
reveals that the sum of the coefficients of a vanishing linear
combination is nonzero. Hence, these coefficients can be chosen so
that they add up to $1$. The corresponding linear combination is
$$
 \textstyle
\frac34 A_1 + \frac38 (A_2 + A_4) - \frac12 A_4 = 0\,.
$$
This yields the relation
 \beq\label{G-qrel}
 \textstyle
\hat f_2
 -
 \left(
\frac38 \hat f_0^2 \hat f_2
 -
\frac12 \hat f_0 \hat f_2 \hat f_0
 +
\frac38 \hat f_2 \hat f_0^2
 \right)
 +
\frac34 \sqrt{4-\hat f_0^2} \, \hat f_2 \, \sqrt{4-\hat f_0^2}
 -
4 \scale^2 \hat f_1^2
 =
0\,.
 \eeq
It holds exactly on the domain of $\hat f_2$ and reproduces the classical
relation \eqref{G-rel} under the na\"ive replacement of the operators $\hat f_i$
by the phase space functions $f_i$, $i=0,1,2$.

From the above relations we can derive
relations, valid on the whole of $\mr D(\hat f_2)$, expressing the operators
$(4-\hat f_0^2) \hat f_2$, $(4-\hat f_0^2)^{\frac12} \, \hat f_2 \, (4-\hat
f_0^2)^{\frac12}$ or $\hat f_2(4-\hat f_0^2)$ as  polynomials in $\hat
f_0$, $\hat f_1$ and the identity. In any of these expressions, the contribution
of the identity is nonzero. Since $(4-\hat f_0^2)$ is given by multiplication by
$4\sinfnpot 2 x$, any subspace on which such a relation can be resolved for
$f_2$ must be contained in
 \beq\label{G-nocore}
 \textstyle
\{\psi\in\mr D(\hat f_2) : \frac{\psi(x)}{\sinfnpot 2 x} \in L^2[0,\pi]\}\,.
 \eeq

\bpr
\label{P-nocore}
The subspace \eqref{G-nocore} is not a core for $\hat f_2$.

\epr

Thus, none of the above relations determines
$\hat f_2$ completely in terms of $\hat f_0$ and $\hat f_1$. 
\\

{\it Proof.}~ Let $D_0$ denote the subspace \eqref{G-nocore}. Let $A$ be defined
as the restriction of $\hat f_2$ to the domain $D_0$. First, we show
that any $\psi\in D_0$ satisfies $\psi'(0) = \psi'(\pi) = 0$. Indeed, since
$\psi'\in\AC^1[0,\pi]$, $\psi'$ is continuous and can be extended
continuously outside $[0,\pi]$. Hence, $\psi'\in C^1[0,\pi]$. Consider the
function $\frac{\psi(x)}{\sinfn x}$ on $]0,\pi[$. Since for $x\to +0$,
$\frac{\psi'(x)}{\cosfn x} \to \psi'(0)$, by the rule of de l'Hospital,
$\frac{\psi(x)}{\sinfn x} \to \psi'(0)$. Hence there exists $x_0>0$ such that
for any $x\in[0,x_0]$ there holds
$\left|\frac{\psi(x)}{\sinfn x}\right| \geq \left|\frac{\psi'(0)}{2}\right|$.
Since $\frac{\psi(x)}{\sinfnpot 2 x}$ is square-integrable then $\psi'(0) = 0$.
A similar argument shows the assertion for $\psi'(\pi)$. Now, integration by
parts yields that for any $\vp\in\AC^2[0,\pi]$ and any $\psi\in D_0$ there holds
$\braket{\vp}{\hat f_2\psi} = \braket{-\hbar^2\beta^2(\vp''+\vp)}{\psi}$. It
follows that $\AC^2[0,\pi] \subseteq \mr D(A^\dagger)$, hence $D_0$ is not a
core for $\hat f_2$, as asserted.
 \qed

 \todo{

-- Gerds Korrekturen

-- Alexanders Korrekturen

-- Erzeugung von $\hat f_2$

Problem: Schreibt man dieselbe Operatorrelation anders (d. h. mit anderen
Polynomen in den $\hat f_i$) auf, dann liefert sie mglw eine andere klassische
Relation, die gar nicht gelten muss. (Pruefe, ob diese Gefahr wirklich besteht.)

Evtl.\ Beispiel angeben, dass naive Ersetzung inkonsistent ist.
 }%


\subsection{Algebra of quantum observables}


As the algebra of quantum observables we would like to take an
algebra which is generated, in some natural way, by the quantized
generators of the algebra of classical observables (to be precise,
the subalgebra of observables polynomial in the position
and momentum variables). There is a natural choice for that
algebra. It relies on the notion of a $C^\ast$-algebra generated
by unbounded operators in the sense of Woronowicz.

\begin{Definition}{\rm \cite{Woro}}\label{D-Woro}

Let $\Hi$ be a separable Hilbert space. Let $\mc A$ be a $C^\ast$-subalgebra of
$\mr B(\Hi)$ and let $T_1,\dots,T_N$ be closed, densely defined operators on
$\Hi$ affiliated with $\mc A$. Then $\mc A$ is generated by $T_1,\dots,T_N$ in
the sense of Woronowicz if for all nondegenerate representations $\pi$ of $\mc
A$ on $\Hi$ and all nondegenerate $C^\ast$-subalgebras $\mc B\subseteq\mr
B(\Hi)$ there holds: if $\pi(T_1),\dots,\pi(T_N)$ are affiliated with $\mc B$,
then $\pi(\mc A)\subseteq \mr M(\mc B)$ and $\pi(\mc A)\mc B$ is dense in $\mc
B$.

\end{Definition}

Here,
$
\mr M(\mc B) = \{ b\in\mr B(\Hi) ~:~ b\mc B, \mc Bb \subseteq \mc
B \}
$
is the multiplier algebra of $\mc B$.
Let us recall the notions entering this definition, see \cite{Woro}.
For a closed, densely defined operator $T$ on $\Hi$, the $z$-transform is
defined by
$$
z_T = T(\II + T^\dagger T)^{-\frac12}\,.
$$
This is a bounded operator on $\Hi$. $T$ can be recovered from $z_T$ by
 \beq\label{G-TfromzT}
T = z_T(\II-z_T^\dagger z_T)^{-\frac12}\,.
 \eeq
A closed, densely defined operator $T$ on $\Hi$ is said to be affiliated with
$\mc A\subseteq\mr B(\Hi)$ (in the sense of Baaj and Julg \cite{BaajJulg})
if $z_T\in\mr M(\mc A)$,
$\II - z_T^\dagger z_T\geq 0$ and $(\II - z_T^\dagger z_T)\mc A$ is dense
in $\mc A$. A nondegenerate representation of $\mc A$ is a $\ast$-morphism
$\pi:\mc A\to\mr B(\Hi)$ such that $\pi(\mc A)\Hi$ is dense in $\Hi$. (Note that
the assumption in Definition \rref{D-Woro} that $\pi(\mc A)$ is dense in
$\mc B$ may not follow from the assumption that $\vp(\mc A)\Hi$ is dense in
$\Hi$ made here.) The representation $\pi:\mc A\to\mr B(\Hi)$ can be extended to
affiliated operators $T$ by extending $\pi$ to $\mr M(\mc A)$ through
$$
\pi(b)\pi(a)\psi = \pi(ba)\psi
 \,,~~~~~~
b\in \mr M(\mc A) \,,~ a\in\mc A\,,~ \psi\in\Hi\,,
$$
and defining $\pi(T)$ by $z_{\pi(T)} = \pi(z_T)$. This definition makes sense,
because $\pi(\mc A)\Hi$ is dense in $\Hi$.

The fundamental criterion to test whether $\mc A$ is generated by a given set of
affiliated operators is

\btm{\rm\cite[Thm.\ 3.3]{Woro}} \label{T-Woro-1}

Let $\mc A$ be a $C^\ast$-subalgebra of $\mr B(\Hi)$ and let
$T_1,\dots,T_N$ be closed, densely defined operators on $\Hi$
affiliated with $\mc A$. Then $\mc A$ is generated by
$T_1,\dots,T_N$ if

{\rm 1.}~ some product built from $(\II + T_i^\dagger T_i)^{-1}$,
$(\II + T_i T_i^\dagger)^{-1}$ belongs to $\mc A$,

{\rm 2.}~ $T_1,\dots,T_N$ separate the representations of $\mc A$
on $\Hi$.
 \qed

\etm

Separation of representations means that for any two distinct
representations $\pi_1,\pi_2 : \mc A \to \mr B(\Hi)$ there exists an 
$i\in\{1,\dots,N\}$ such that $\pi_1(T_i)\neq \pi_2(T_i)$.
Prominent examples of $C^\ast$-algebras generated by unbounded
operators are:
\medskip

1.~ Let $\mc A$ be a unital $C^\ast$-subalgebra of $\mr B(\Hi)$
and let $T_1,\dots,T_N$ be elements of $\mc A$ such that the
algebra generated by $T_1,\dots,T_N$ and the identity is dense in $\mc A$. Then
$T_1,\dots,T_N$ generate $\mc A$ in the sense of Woronowicz. Thus, the concept
is a generalization of the ordinary notion of generation of an
algebra by a subset.
\medskip

2.~ Let $G$ be a connected Lie group, let $\Hi = L^2(G)$ and let
$\mc A$ be the group $C^\ast$-algebra $C^\ast(G)$. Then $\mc A$ is
generated in the sense of Woronowicz by any basis in the Lie
algebra $\mf g$, where the basis elements are viewed as first
order differential operators on $G$.
 \medskip

3.~ Let $\Hi = L^2(\RR)$, let $\mc A$ be the $C^\ast$-algebra $\mr
K(\Hi)$ of compact operators on $\Hi$. Then $\mc A$ is generated
in the sense of Woronowicz by the position operator $T_1 = x$ and
the momentum operator $T_2 = \frac\hbar{\mr i} \frac{\mr d}{\mr
dx}$.
 \medskip

 \bre

At the present stage this theory has the disadvantage of not providing a general method
how to construct the algebra for a given set of generators.

 \ere

\btm\label{T-Woro}

Each of the sets $\{\hat f_0, \hat f_2\}$ and $\{\hat f_1, \hat
f_2\}$ generates the $C^\ast$-algebra $\mr K\big(L^2(G)^G\big)$ in
the sense of Woronowicz.

\etm

Thus, it is natural to define the algebra of quantum observables
to be
$$
\mc O_q := \mr K\big(L^2(G)^G\big)\,.
$$

{\it Proof.}~ In the proof, denote $\Hi = L^2(G)^G$. The $\hat
f_i$ are affiliated with $\mr K(\Hi)$, because this holds for any
closed and densely defined operator. To prove Theorem
\rref{T-Woro}, we use the criterion given in Theorem
\rref{T-Woro-1}. The first condition holds, because $\hat f_2^2$
has eigenbasis $\chi_n$ with eigenvalues $\hbar^4 \beta^4
n^2(n+2)^2$ and hence has compact resolvent. To check the second
condition, let $k=0$ or $1$. Assume that we are given
representations $\pi_1$, $\pi_2$ with $\pi_1(\hat f_i) =
\pi_2(\hat f_i)$, $i=k,2$. By definition of the operators
$\pi_j(\hat f_i)$, then $\pi_1(z_{\hat f_i}) = \pi_2(z_{\hat
f_i})$, $i=k,2$. Hence, $\pi_1$ and $\pi_2$ coincide on the
subalgebra $\tilde A$ of the multiplier algebra of $\mr K(\Hi)$
generated by $\II$, $z_{\hat f_k}$, $z_{\hat f_2}$. To prove
$\pi_1 = \pi_2$ it suffices to show $\mr K(\Hi) \subseteq
\tilde{\mc A}$. As the multiplier algebra of $\mr K(\Hi)$ is $\mr
B(\Hi)$, we can apply the following criterion.

\ble\cite[Prop.\ 10.4.1]{KaRi}\label{L-Woro-2}

Let $\tilde{\mc A}$ be a $C^\ast$-subalgebra of $\mr B(\Hi)$. If

{\rm 1.}~ $\tilde{\mc A} \cap \mr K(\Hi) \neq 0$,

{\rm 2.}~ $\tilde{\mc A}$ is irreducibly represented on $\Hi$,

then $\mr K(\Hi) \subseteq \tilde{\mc A}$.
 \qed

\ele

We check these two conditions. For the first one, we use that
for a closed, densely defined operator $T$ there holds the
identity
 \begin{align*}
\II - z_T^\dagger z_T
 & =
(\II + T^\dagger T)^{-1}\,,
 \end{align*}
see e.g.\ \cite{Lance}. Plugging in $\hat f_2$ for $T$ we observe: the
l.h.s.\ belongs to $\tilde{\mc A}$ and the r.h.s.\ was shown above
to belong to $\mr K(\Hi)$. Thus, the first condition holds,
indeed. To prove the second condition, we apply the lemma of
Schur. Assume that we are given a bounded operator $S$ on $\Hi$
that commutes with all elements of $\tilde{\mc A}$. Then $S$
commutes with $z_{\hat f_k}$ and $z_{\hat f_2}$. In particular, $S$ leaves
invariant the eigenspaces of $z_{\hat f_2}$. According to \eqref{G-qobs-bs-2},
$\chi_n$ is a basis of eigenvectors of $z_{\hat f_2}$ with eigenvalues
$\frac{\hbar^2\beta^2 n(n+2)}{\sqrt{1+\hbar^4\beta^4 n^2(n+2)^2}}$. Since this
is a strictly monotonous function of $n$, the eigenspaces have dimension $1$.
Hence, $S\chi_n = \lambda_n \chi_n$, $n=0,1,2,\dots$ with
$\lambda_n\in\CC$. According to \eqref{G-TfromzT}, $[S,z_{\hat f_k}] = 0$
implies $S\hat f_k \chi_n - \hat f_k S \chi_n = 0$ for all $n$. This yields,
respectively,
 \begin{align*}
 \textstyle
(\lambda_{n+1} - \lambda_n) \chi_{n+1}
 +
(\lambda_{n-1} - \lambda_n) \chi_{n-1}
 & = 0 & & (k=0)
\\
 \textstyle
\frac{2n+3}{4} \left(\lambda_{n+1} - \lambda_n\right) \chi_{n+1}
 -
\frac{2n+1}{4} \left(\lambda_{n-1} - \lambda_n\right) \chi_{n-1}
 & = 0 & & (k=1)
 \end{align*}
for all $n$. In both cases, it follows $\lambda_{n+1} = \lambda_n$ for all $n$,
hence $S = \lambda \II$. Then $\tilde{\mc A}$ is irreducibly represented on
$\Hi$ by the lemma of Schur and hence the second condition of Lemma
\rref{L-Woro-2} is satisfied. This shows that condition 2 of Theorem
\rref{T-Woro-1} holds and, therefore, completes the proof of Theorem
\rref{T-Woro}.
 \qed

\bre

There remains the question whether the set $\{\hat f_0,\hat f_1\}$ generates
$\mr K\big(L^2(G)^G\big)$  as well. The crucial point
is Condition 1 of Theorem \rref{T-Woro-1}. While according to Proposition
\rref{P-spec}, the operators $(\II+\hat f_0^2)^{-1}$ and $(\II+\hat
f_1^2)^{-1}$ do not belong to $\mr K(L^2(G)^G)$, it would be sufficient to show that some product of these 
operators is compact. We did not succeed to clarify this point.
 \todo{

\RA~ pruefe, ob nicht doch ein Produkt von $(\II + \hat
f_0^2)^{-1}$ und $(\II + \hat f_1^2)^{-1}$ kompakt ist.

 }

\ere


\subsection{Relation with the algebra of bosonic quantum
observables in lattice gauge theory}


We discuss the relation between the algebra of quantum observables
of our model and the bosonic part of the algebra of observables of
a quantum lattice gauge theory of \cite{KiRu05}. This paper
concentrates on the case of lattice quantum chromodynamics, i.e.,
gauge group $\SU(3)$ rather than $\SU(2)$ as in our model.
However, the results are valid for general $\SU(n)$. We recall the
construction of the bosonic observable algebra for the case of a
single plaquette without external links, after having implemented
the tree gauge. The bosonic field algebra is the crossed product
algebra $\mc F = C(G)\otimes_\alpha G$ associated with the
$C^\ast$-dynamical system $(C(G),G,(\mr L_{g^{-1}})^\ast)$. For
the notions of $C^\ast$-dynamical system and crossed-product
algebra, see \cite{Pedersen}.
$\mc F$ carries a natural $G$-action. The bosonic observable
algebra $\mc O$ is defined as the quotient of the subalgebra of
$G$-invariant elements of $\mc F$ by the ideal defined by the
generators of the $G$-action. This factorization corresponds to
imposing the Gauss law, which is the quantum analogue of the
restriction of the phase space to the zero level set of the
momentum mapping. If there are external links, this definition of
$\mc O$ yields the subalgebra of internal observables. The natural
covariant representation of $(C(G),G,\alpha)$ on $L^2(G)$
naturally induces a representation of $\mc F$ on $L^2(G)$, mapping
$\mc F$ to $\mr K(L^2(G))$.
 \todo{

Evtl. Definition der kovar Dst und der induzierten von $\mc F$
angeben. Erstere: $\pi(f)\psi = f\psi$, $u(g)\psi = \mr
(L_{g^{-1}})^\ast\psi$.

 }%
It is shown in \cite{KiRu05} that this representation is the
unique irreducible representation of $\mc F$. It is therefore
called the generalized Schr\"odinger representation. Using this
representation, it is then shown that $\mc O$ can be identified
with the compact operators on the closed subspace $L^2(G)^G$.
Thus, through this identification, the algebra of quantum
observables $\mc O_q$ of our model coincides with the bosonic
observable algebra $\mc O$ of \cite{KiRu05}, specified to the case
of a single plaquette without external links.

Next, we compare generators. Let $U^A{}_B : G\to\CC$ denote the matrix entry
functions. Choose a basis $T_i$ in $\mf g$ orthonormal
w.r.t.\ the trace form and define vector fields $E^A{}_B$ on $G$
by
$$
 \textstyle
E^A{}_B = \sum_i (T_i)^A{}_B T_i\,,
$$
where $(T_i)^A{}_B$ are the entries of the basis element $T_i$
when viewed as a matrix, whereas the second $T_i$ is viewed as a
vector field. It is stated in \cite{KiRu05} that $\mc F$, when
realized as $\mr K(L^2(G))$, is generated in the sense of
Woronowicz by the multiplication operators $U^A{}_B$ and the first
order differential operators $E^A{}_B$. It was not clarified in \cite{KiRu05} whether gauge invariant 
combinations of $U^A{}_B$ and $E^A{}_B$ generate the observable algebra. 

Our quantum observables $\hat f_0$, $\hat f_1$ and $\hat f_2$ can be expressed
in terms of the gauge invariant combinations $U^A{}_A\, , $ $U^A{}_B E^B{}_A$ and $E^A{}_B E^B{}_A$ as follows. 
Since $U^A{}_A = f_0$ as functions
on $G$, for the multiplication operators we have
$$
\hat f_0 = U^A{}_A\,.
$$
For the value of the vector field $U^A{}_B E^B{}_A$ at
$a\in G$ we find
$\left(U^A{}_B E^B{}_A\right)_a
 =
\mr L_a' \sum_i \tr(a T_i) T_i$. Since $T_i$ is orthonormal w.r.t.\ the trace
form, $\sum_i \tr(a T_i) T_i = -\OPg(a)$. According to \eqref{G-Yi}, then
$U^A{}_B E^B{}_A = - Y_{f_1}$ and hence
$$
 \textstyle
\hat f_1
 =
\mr i\hbar\big( - U^A{}_B E^B{}_A + \frac34 U^A{}_A\big)\,.
$$
Finally, a similar computation shows
$$
\hat f_2 = 2\hbar^2\beta^2 E^A{}_B E^B{}_A\,.
$$
Thus, Theorem \rref{T-Woro} implies that, in the case of a single plaquette
without external links, the algebra of quantum observables of \cite{KiRu05} is
generated, in the sense of Woronowicz, by $U^A{}_A$ and $E^A{}_B E^B{}_A$ or by
$U^A{}_B E^B{}_A$ and $E^A{}_B E^B{}_A$. This
extends the result of \cite{KiRu05} on the generation of the field algebra by
unbounded operators  to the algebra of observables, at
least in the simple case at hand.

In \cite{KiRu05} it was argued that on a purely algebraic level the observable algebra is 
generated by $U^A{}_A$ and $U^A{}_B E^B{}_A$ and that all other invariants can be 
expressed in terms of these generators. This is, however, the pair of operators for 
which we could not prove that they generate the algebra in the sense of Woronowicz. Moreover, from Proposition \ref{P-nocore} 
we conclude that e.g. the quadratic Casimir operator $E^A{}_B E^B{}_A$ cannot be expressed in 
terms of $U^A{}_A$ and $U^A{}_B E^B{}_A$ on a core. These observations show that any na\"ive algebraic procedure of 
reducing the number of independent generators has to be handled with care.


\subsection{Towards quantum dynamics}


Quantization of the classical Hamiltonian \eqref{G-Ham-ivr} yields the quantum
Hamiltonian
$$
 \textstyle
\hat H = \frac12 \hat f_2 + \frac{1}{2g^2}(3-\hat f_0)
$$
which is a time-independent self-adjoint operator with domain $\mr D(\hat H) =
\mr D(\hat f_2)$. On the level of pure states (Schr\"odinger picture),
dynamics is given by the 1-parameter group of unitary transformations of
$L^2(G)^G$ generated by $\hat H$,
$$
U_t = \mr e^{-\frac{\mr i}{\hbar} \hat H t}\,.
$$
Since the algebra of compact operators is invariant under unitary
tranformations, $U_t$ induces a 1-parameter automorphisms group $\alpha_t$ of
the algebra of quantum observables by
$$
\alpha_t(A) = U_t A U_t^\dagger\,.
$$
On the level of observables (Heisenberg picture), dynamics is given by the
1-parameter automorphism 
group $\alpha_t$. It is interesting as well to study the dynamics of the
generators $\hat f_k$. On the common invariant core $C^\infty(G)^G$ it is given
by 
\beq
\label{G-Hbeq0}
\hat f_k(t) = U_t \hat f_k U_t^\dagger \, .
\eeq
The corresponding equation of motion, on this core, reads
 \beq
 \label{G-Hbeq}
\ddt \hat f_k(t) = \frac{\mr i}{\hbar} [\hat H,\hat f_k(t)]
 \,,~~~~~~
\hat f_k(0) = \hat f_k
 \,,~~~~~~
k=0,1,2\,.
 \eeq
The automorphism group $\alpha_t$ and the operators $\hat f_k(t)$ will be
studied elsewhere. 

We conclude with a discussion of the commutators between the
generators $\hat f_k \, . $ These commutators are relevant for the evaluation of the right-hand side of \eqref{G-Hbeq0} 
and for the iterative solution of \eqref{G-Hbeq}, respectively. Since $\hat f_0$ leaves invariant the domains of
$\hat f_1$ and 
$\hat f_2$, the commutators $[\hat f_0,\hat f_1]$ and $[\hat f_0,\hat f_2]$ are
defined on these domains. A straightforward computation using
\eqref{G-f0expr}--\eqref{G-f2expr} yields 
 \beq\label{G-ctr-f0}
[\hat f_0,\hat f_1] = \mr i \hbar \left(2 - \frac12\hat
f_0^2\right)
 \,,~~~~~~
[\hat f_0,\hat f_2] = 4 \scale^2 \, \mr i \hbar \, \hat f_1\,.
 \eeq
We claim that the commutator of $\hat f_1$ and $\hat f_2$ is
defined on $\mr D(\hat f_2)$ and is given by
 \beq\label{G-ctr-f1f2}
 \textstyle
[\hat f_1,\hat f_2]
 =
- \frac12 \mr i \hbar
 \left(
\hat f_0 \hat f_2 + \hat f_2 \hat f_0 + 3 \hbar^2 \hat f_0
 \right)\,.
 \eeq
To see this, write
 \begin{align*}
\hat f_1 \hat f_2 - \hat f_2 \hat f_1
 & =
 \textstyle
-\mr i \hbar^3 \beta^2
 \left\{
 \left(
\ddx\sinfn x - \frac12\cosfn x
 \right)
 \big(
\ddxx + 1
 \big)
 -
 \big(
\ddxx + 1
 \big)
 \left(
\ddx\sinfn x - \frac12\cosfn x
 \right)
 \right\}
\\
 & =
 \textstyle
\mr i \hbar^3 \beta^2
 \left\{
\ddx
 \big(
\ddxx\sinfn x - \sinfn x\ddxx
 \big)
 +
\frac12
 \big(
\cosfn x \ddxx - \ddxx \cosfn x
 \big)
 \right\}
 \end{align*}
and observe that for $\psi\in\AC^2[0,\pi]$,
$$
 \textstyle
\left(\ddxx\sinfn x - \sinfn x\ddxx\right) \psi(x)
 =
\left(-\sinfn x + 2\cosfn x \ddx\right) \psi(x)\,.
$$
Hence $\hat f_1\hat f_2 - \hat f_2 \hat f_1$ contains derivatives
up to second order only and is therefore defined on $\mr D(\hat f_2) \subseteq
\AC^2[0,\pi]$, indeed. Then a straighforward calculation yields
\eqref{G-ctr-f1f2}. Thus, all the commutators between the quantum observables
$\hat f_k$ are defined on $\mr D(\hat f_2)$.

\bre

Comparing the commutators \eqref{G-ctr-f0} and \eqref{G-ctr-f1f2} with the
corresponding Poisson brackets \eqref{G-Poibra} we observe
that for the combinations of $\hat f_0$ with $\hat f_1$ and $\hat f_2$ the
na\"ive relation between the commutator and the Poisson bracket (replacing, in
the commutator, the operators by their classical counterparts,
provided the latter are well-defined) holds exactly. For the
combination of $\hat f_1$ and $\hat f_2$, this relation holds in the limit
$\hbar \to 0$.

\ere


\section{Outlook}


An obvious future task is to generalize the results of this paper
to a general compact Lie group. Another task is to study how the
algebra of quantum observables depends on the choice of what phase
space functions should be considered polynomial, cf.\ Remark
\rref{Rem-pnmalg}. E.g., one should carry out a similar
construction for the generators of the algebra of real invariant
polynomials on $G^\CC$ and compare the resulting algebra of
quantum observables with the one obtained above. The concept used in this paper
should be also compared with an alternative approach proposed by Buchholz and Grundling 
\cite{BuchholzGrundling}, who define the notion
of resolvent algebra associated with a symplectic space and
propose to take this algebra as the algebra of observables in a
bosonic field theory. The resolvent algebra is a unital
$C^\ast$-algebra defined abstractly in terms of generators and
relations. Equivalently, it can be viewed as  generated by
the resolvents $(\mr i\lambda - \phi(f))^{-1}$ of the field
operators $\phi(f)$ of some quantum field $\phi$.
Following these ideas, in our model one may take the unital
$C^\ast$-algebra generated by the resolvents $(\mr i\lambda - \hat
f_k)^{-1}$ of the quantized generators $\hat f_0$, $\hat f_1$,
$\hat f_2$ as the algebra of observables. This algebra will be
studied elsewhere. It is definitely distinct from the algebra of
quantum observables $\mc O_q$ constructed above. The trivial
reason for that is that this algebra is unital by definition;
another reason is that, according to Proposition \rref{P-spec},
the resolvents $(\mr i\lambda - \hat f_0)^{-1}$ and $(\mr i\lambda
- \hat f_1)^{-1}$ are not compact.

Furthermore, we address the problem of studying the influence of the
stratification of the classical configuration and phase spaces on the quantum
theory. For that purpose, one has to find a quantum structure that implements
this stratification. On the level of pure states, such a quantum structure is
given by a costratification of the Hilbert space \cite{Hue:qr}. One may think of a
costratification as a family of closed subspaces, indexed by the
strata, and a family of orthoprojectors, indexed by the inclusion relations
between the closures of the strata. For the model under consideration, the
costratified Hilbert space was studied in \cite{hrs}. On the other hand, it is
not clear how to implement the stratification on the level of observables.
For the concrete algebra of observables at hand, this problem will be studied in
detail in a future work.

\section{Acknowledgements}

The authors are grateful to A.\ Hertsch, J.\ Huebschmann, J.\ Kijowski and
K.\ Schm\"udgen for helpful discussions and to K.\ Schm\"udgen for reading part
of the manuscript.

\begin{appendix}

\section*{Appendix}

We prove Formula \eqref{G-f1-4}. Choose an orthonormal basis $B_i$ in $\mf g$.
Let $\beta_i$ denote the
elements of the dual basis in $\mf g^\ast$. Then $\beta_i(A) =
\langle B_i,A \rangle$ for any $A\in\mf g$. We have $v =
\beta_1\wedge\beta_2\wedge\beta_3$, where the $\beta_i$ are viewed
as left-invariant forms. Using the derivation property of the Lie
derivative, expanding
$$
 \textstyle
\mc L_{Y_{f_1}} \beta_i
 =
\sum\nolimits_{j=1}^3 ~ (\mc L_{Y_{f_1}} \beta_i) (B_j) ~ \beta_j
$$
and rewriting $(\mc L_{Y_{f_1}} \beta_i) (B_i) = - \beta_i (\mc
L_{Y_{f_1}} B_i) = - \langle B_i,\mc L_{Y_{f_1}} B_i \rangle$ we
obtain
 \beq\label{G-gf1}
 \textstyle
\mc L_{Y_{f_1}} v
 =
-
 \left(
\sum\nolimits_{i=1}^3 ~ \langle B_i,\mc L_{Y_{f_1}} B_i\rangle
 \right)
v\,.
 \eeq
We calculate $\mc L_{Y_{f_1}} B_i$ by taking derivatives in the
ambient vector space $\mr M_2(\CC)$. According to \eqref{G-Yi}, for $a\in G$,
$$
 \textstyle
(\mc L_{Y_{f_1}} B_i)_a
 =
[Y_{f_1},B_i]_a
 =
\ddtn \ddsn a \mr e^{\OPg(a)t} \mr e^{B_i s}
 -
\ddtn \ddsn a \mr e^{B_i t} \mr e^{\OPg(a\mr e^{B_it})s}\,.
$$
This yields $a \OPg(a) B_i - a B_i \OPg(a) - a \OPg(aB_i)$, which
can be rewritten as $- a \OPg(B_i a)$. Hence,
$$
(\mc L_{Y_{f_1}} B_i)_a = - \mr L_a' \OPg(B_i a)\,.
$$
Then
$$
 \textstyle
\langle B_i,\mc L_{Y_{f_1}} B_i\rangle(a)
 =
-\langle B_i,\OPg(B_ia) \rangle
 =
\frac{1}{2\scale^2}
 ~\frac12~
\tr\left(B_i^2 (a + a^\dagger) \right)
 =
\frac{1}{2\scale^2} ~\frac12~ \tr(B_i^2) \tr(a)
 =
- \frac12\tr(a)\,,
$$
where we have used $B_i^2 -\frac12\tr(B_i^2)\II = 0$, due to the
Cayley-Hamilton theorem. Then \eqref{G-gf1} yields $\mc L_{Y_{f_1}} v =
\frac{3}{2} f_0 \, v$, i.e., Formula \eqref{G-f1-4}.

\end{appendix}


\begin{thebibliography}{99}

\onehalfspacing

\bibitem{ACG}
 J.M.\ Arms, R.H.\ Cushman, M.J.\ Gotay,
 in {\it The geometry of Hamiltonian systems},
 Math.\ Sci.\ Res.\ Inst.\ Publ.\ 22, pp.\ 33--51 (Springer, New York, 1991).


\bibitem{BaajJulg}
 S.\ Baaj, P.\ Julg,
 C.\ R.\ Acad.\ Sci.\ Paris, Sér. I Math., {\bf 296}, 875--878 (1983).

\bibitem{Buchholz82}
 D.\ Buchholz,
 Commun.\ Math.\ Phys.\ {\bf 85}, 49--71 (1982).


\bibitem{Buchholz86}
 D.\ Buchholz,
 Phys.\ Lett. B {\bf 174}, 331--334 (1986).


\bibitem{BuchholzGrundling}
 D.\ Buchholz, H.\ Grundling,
 arXiv:0705.1988v3; to appear in J.\ Funct.\ Anal.

\bibitem{DHR71}
 S.\ Doplicher, R.\ Haag, J.\ Roberts,
 Commun.\ Math.\ Phys.\ {\bf 23},  199--230 (1971).


\bibitem{DHR74}
 S.\ Doplicher, R.\ Haag, J.\ Roberts,
 Commun.\ Math.\ Phys.\ {\bf 35},  49--85 (1974).


\bibitem{DR90}
 S.\ Doplicher, J.\ Roberts,
 Commun.\ Math.\ Phys.\ {\bf 131}, 51--107 (1990).


\bibitem{Fredenhagen}
 K.\ Fredenhagen, M.\ Marcu,
 Commun.\ Math.\ Phys.\ {\bf 92}, 81--119 (1983).


\bibitem{Froehlich}
 J.\ Fr\"ohlich,
 Commun.\ Math.\ Phys.\ {\bf 66}, 223--265 (1979).


\bibitem{Hall:Compact}
 B.C.\ Hall,
 Commun.\ Math.\ Phys.\ {\bf 226}, 233--268 (2002).


\bibitem{Helgason}
 S.\ Helgason,
 {\em Groups and geometric analysis. Integral geometry, invariant differential
 operators, and spherical functions} (Academic Press 1984).


\bibitem{Hue:qr}
 J.\ Huebschmann,
 J.\ reine angew.\ Math.\ \textbf{591}, 75--109 (2006).


\bibitem{Hue:bedlewo}
 J.\ Huebschmann,
 in {\em The mathematical legacy of C.\ Ehresmann}, pp.\  325--347 
 (Banach Center Publications 76, Warsaw 2007).


\bibitem{Hue:holopewe}
 J.\ Huebschmann,
 J.\ Geom.\ Phys.\ 58, 833--848 (2008).


\bibitem{hrs}
 J.\ Huebschmann, G.\ Rudolph, M.\ Schmidt,
 hep-th/0702017.
\\
 J.\ Huebschmann, G.\ Rudolph, M.\ Schmidt,
 in {\it Lie Theory and its Applications in
 Physics}, pp.\ 190--210 (Heron Press, Sofia, 2008).

\bibitem{JaKiRu}
 P.D.\ Jarvis, J.\ Kijowski, G.\ Rudolph,
 J.\ Phys.\ A {\bf 38}, 5359--5377 (2005).


\bibitem{KaRi} 
 R.V.\ Kadison, J.R.\ Ringrose,
 {\em Fundamentals of the Theory of Operator Algebras}
 (Academic Press 1986).


\bibitem{KiRu02}
 J.\ Kijowski, G.\ Rudolph,
 J.\ Math.\ Phys.\ {\bf 43}, 1796--1808 (2002).


\bibitem{KiRu05}
 J.\ Kijowski, G.\ Rudolph,
 J.\ Math.\ Phys.\ {\bf 46}, 032303 (2005).


\bibitem{KiRuSl}
 J.\ Kijowski, G.\ Rudolph, C.\ \'Sliwa,
 Lett.\ Math.\ Phys.\ {\bf 43}, 299--308 (1998).


\bibitem{KiRuTh}
 J.\ Kijowski, G.\ Rudolph, A.\ Thielmann,
 Commun.\ Math.\ Phys.\ {\bf 188}, 535--564 (1997).


\bibitem{Lance}
 E.\ Lance,
 {\it Hilbert $C^\ast$-modules}
 (Cambridge University Press, Cambridge, 1995).


\bibitem{Pedersen}
 G.K.\ Pedersen,
 {\em $C\sp{*} $-algebras and their automorphism groups}
 (Academic Press, Inc., London-New York, 1979).


\bibitem{rsv:review}
 G.\ Rudolph, M.\ Schmidt, I.P.\ Volobuev,
 J.\ Phys.\ A: Math.\ Gen.\ {\bf 35}, R1--R50 (2002).

\bibitem{Schwarz}
 G.W.\ Schwarz,
 Topology {\bf 14}, 63--68 (1975).


\bibitem{Woodhouse}
 J.\ \'Sniatycki,
 {\em Geometric quantization and Quantum Mechanics}
 (Springer, 1980).
\\
 N.M.J.\ Woodhouse,
 {\em Geometric quantization}
 (Clarendon Press, Oxford 1991).


\bibitem{StrocchiWightman}
 F.\ Strocchi, A.\ Wightman,
 J.\ Math.\ Phys.\ {\bf 15}, 2198--2224 (1974).
 \\
 F.\ Strocchi,
 Phys.\ Rev.\ D {\bf 17}, 2010--2021 (1978).


\bibitem{Weyl}
 H.\ Weyl,
 {\it The classical groups}
 (Princeton University Press, Princeton, N.J., 1946).


\bibitem{Woro}
 S.L.\ Woronowicz,
 Rev.\ Math.\ Phys.\ {\bf 7}, 481--521 (1995).


\bibitem{Wren}
 K.K.\ Wren,
 J.\ Geom.\ Phys.\ {\bf 24}, 173--202 (1998).
 \\
 K.K.\ Wren,
 Nucl.\ Phys.\ {\bf B 521}, 471--502 (1998).




\end{thebibliography}
\end{document}